\pdfoutput=1
\documentclass[10pt,twocolumn,superscriptaddress,preprintnumbers,nofootinbib,amsmath,amssymb,aps,prl]{revtex4-2}
\usepackage{mathrsfs}
\usepackage{booktabs}
\usepackage[normalem]{ulem}

\usepackage{tikz}
\usetikzlibrary{calc,decorations.markings,arrows.meta,positioning,backgrounds}

\usepackage{hyperref}
\hypersetup{pdftitle={},pdfcreator={},linkcolor=[rgb]{0.15,0.35,0.75},colorlinks=true,citecolor=[rgb]{0.675,0,0.2},urlcolor=[rgb]{0.15,0.35,0.65}}

\newcommand{\od}{\mathrm{d}}
\newcommand{\ireg}[2]{\int_{#1,\mathrm{reg}}^{#2}}
\newcommand{\ldeg}{\operatorname{ldeg}}
\newcommand{\BNP}{\mathcal{B}_{\mathrm{NP}}}

\allowdisplaybreaks

\begin{document}

\preprint{DESY 26-171}

\title{Pushing Feynman integrals to higher loops via direct parametric integration}

\author{L.~Cesinger}
\email{lucas.cesinger@desy.de}
\affiliation{Deutsches Elektronen-Synchrotron DESY, Notkestr.~85, 22607 Hamburg, Germany}
\author{A.~Kardos}
\email{kardos.adam@science.unideb.hu}
\affiliation{
  Institute of Physics,
  University of Debrecen,
  4010 Debrecen, PO Box 105, Hungary}
\affiliation{Institute for Theoretical Physics, ELTE E\"otv\"os Lor\'and University,
  1117 Budapest, Hungary}
\author{M.~Klann}
\email{marco.klann@desy.de}
\author{S.~Moch}
\email{sven-olaf.moch@desy.de}
\affiliation{II. Institute for Theoretical Physics, Hamburg University,
  22761 Hamburg, Germany}
\author{O.~Schnetz}
\email{oliver.schnetz@mi.uni-erlangen.de}
\affiliation{II. Institute for Theoretical Physics, Hamburg University,
  22761 Hamburg, Germany}
\affiliation{Mathematical Institute, University of Oxford, Oxford OX2 6GG, UK}

\begin{abstract}
Parametric integration of Feynman integrals provides a powerful approach for obtaining analytic results in terms of hyperlogarithms.
We apply a novel algorithm to regularise iterated integrals in dimensional regularisation, with $D = 4 - 2\varepsilon$, which enables the direct computation of the coefficients in the Laurent expansion in $\varepsilon$.
We demonstrate the efficiency of the method by computing new results in on-shell kinematics for the massless ladder-box diagram at four and five loops and for the massless non-planar box diagram at four loops.
As a by-product, we conjecture an all-loop generating function for the leading poles of the ladder-box diagrams.
\end{abstract}

\maketitle

Multi-loop Feynman integrals are essential ingredients in the computation of scattering amplitudes in perturbative quantum field theory (QFT) for precision predictions at particle colliders~\cite{Heinrich:2020ybq}.
Recent advances in precision calculations are driven by the increasing accuracy of experimental measurements expected during upcoming runs of the Large Hadron Collider, particularly in the Higgs sector of the Standard Model of particle physics, as well as by the challenges posed by future colliders such as the FCC-ee~\cite{deBlas:2025gyz}.
A wide variety of numerical and analytical methods are available for the computation of Feynman integrals~\cite{Weinzierl:2022eaz} and scattering amplitudes~\cite{Travaglini:2022uwo}.
At higher loop orders, complications arise due to the ubiquitous ultraviolet and infrared singularities of QFT, which are typically treated using dimensional regularisation in $D=4-2\varepsilon$ dimensions~\cite{tHooft:1972tcz,Bollini:1972ui}.

The approach predominantly used in the community follows a relatively standardised workflow.
It starts with the generation of the relevant Feynman diagrams and the identification of integral families associated with the graph topologies under consideration.
Integration-by-parts (IBP) identities are then used to reduce the integrals to a set of so-called master integrals, for which systems of differential equations in the kinematic variables are constructed and solved subject to suitable boundary conditions.
The solutions are typically expressed as Laurent series in $\varepsilon$.
Boundary values are commonly obtained numerically, e.g.\ with \textsc{AMFlow}~\cite{Liu:2022chg}, and converted into analytic expressions using integer-relation algorithms such as PSLQ~\cite{PSLQ} or LLL~\cite{Lenstra:1982eee}.
Strictly speaking, the resulting expressions are therefore numerically fitted rather than derived analytically.
A complementary approach, which is not restricted to multi-scale problems and yields the boundary values exactly, is the direct integration over Feynman parameters.

Parametric integration of Feynman integrals proceeds in three steps: (1) rewrite the Feynman integral in terms of Schwinger parameters using some parametric representation, so that the resulting integral is expressed in terms of the Symanzik graph polynomials, which encode the topology of the diagram and its dependence on the external kinematics; (2) regularise the integral; and (3) choose an integration order and integrate the parameters successively in terms of hyperlogarithms, following Refs.~\cite{Brown:2008um,Brown:2009ta}.
Publicly available implementations in computer algebra systems include \texttt{HyperInt}~\cite{Panzer:2014caa} and \texttt{HyperFORM}~\cite{Kardos:2025klp}.
The third step requires linear reducibility, while the expansion in $\varepsilon$ under the integral sign requires a suitably regularised integrand.
For divergent integrals, the latter condition is not generally satisfied without additional regularisation.

We address this difficulty with a novel regularisation method based on regularised limits at tangential base points, a concept originating in Ref.~\cite{Deligne:1989}.
In contrast to integrand-level regularisation, which can lead to expression swell similar to that encountered in IBP reductions, the method regularises the integral itself.
Each divergence gives rise to a boundary term containing an explicit pole of the form $1/(a\varepsilon)$, while the remainder is a regularised integral whose integrand is naively Taylor expanded in $\varepsilon$ and integrated term by term.
Each extracted boundary term reduces the number of parametric integrations by one, so that higher degrees of divergence lead to expansions to higher order in $\varepsilon$ but fewer integrations.
Overall, the resulting integrals are easier to evaluate.
The mathematical foundations of the method are presented in a companion paper~\cite{Schnetz:2026}.
Other ideas about the regularisation of divergent integrals have recently been put forward in Refs.~\cite{Hillman:2023ezp,Salvatori:2024nva,Giroux:2026tgd}.

A minimal example of the generic situation is
\begin{equation}
  I(\varepsilon) = \int_0^\infty \od x\, \frac{x^{-1+\varepsilon}}{x+1} = \frac{\pi}{\sin \pi\varepsilon} = \frac{1}{\varepsilon} + \zeta_2\,\varepsilon + \mathcal{O}(\varepsilon^3)\,.
  \label{eq:minimal}
\end{equation}
Expansion in $\varepsilon$ and integration do not commute.

The regularised limit of a function $f$ as $x\to 0$ of the form
\begin{equation}
  f(x) = \sum_{n\ge N}\sum_{k=0}^{K} c_{n,k}\, x^n \log^k x\,,\quad N\in\mathbb{Z}\, ,
\end{equation}
is defined as (a special case of the tangential base point prescription in Ref.~\cite{Deligne:1989})
\begin{equation}
  \lim_{x\to0,{\mathrm{reg}}} f(x) = c_{0,0}\,,\quad\text{i.e.}\quad (\log x)^k x^n\big|_{x=0,{\mathrm{reg}}} = \delta_{k,0}\,\delta_{n,0}\,.
  \label{eq:reg_limits}
\end{equation}
For $f$ as above with primitive $F$, the regularised integral is defined as
\begin{equation}
  \ireg{0}{b} \od x\, f(x) = F(b) - \lim_{x\to0,{\mathrm{reg}}} F(x)\, .
  \label{eq:reg_ints}
\end{equation}
An analogous prescription is applied for regularised limits at $\infty$.
This definition is applied to each coefficient of the integrand's Taylor expansion in $\varepsilon$.
The coefficients of $x^{-n+a\varepsilon}f(x)$ are $x^{-n}\{(a\log x)^k/k!\}\, f(x)$ and hence belong to the class considered here if $f$ admits a Laurent expansion at $x=0$.
For such a function, with $x^{-1+a\varepsilon}f(x)$ integrable at $\infty$ and $a\ne0$, one has
\begin{equation}
  I(\varepsilon)
  = \int_0^\infty \od x\, x^{-1+a\varepsilon} f(x)\,
  = I_{\mathrm{reg}}(\varepsilon) + \frac{f(0)}{a\varepsilon}
  \label{eq:ireg-example}
  \, ,
\end{equation}
where the regularised integral $I_{\mathrm{reg}}(\varepsilon)$,
\begin{equation}
  I_{\mathrm{reg}}(\varepsilon) = \ireg{0}{\infty} \od x\, x^{-1+a\varepsilon}f(x)\, ,
\end{equation}
has to be integrated termwise in $\varepsilon$ in a Laurent expansion of the integrand with the rules of Eqs.~\eqref{eq:reg_limits} and \eqref{eq:reg_ints}.
The example in Eq.~\eqref{eq:minimal} is recovered using $f(0) = 1$ and $I_{\mathrm{reg}}(\varepsilon) = \zeta_2\varepsilon + \mathcal{O}(\varepsilon^3)$.

Calculus with regularised integrals has the following properties:
For convergent integrals, it agrees with ordinary integration, $\ireg{0}{\infty} = \int_0^\infty$, and integration by parts remains valid
\begin{equation}
  \ireg{0}{\infty}\od x\, f'g = fg\big|_{0,{\mathrm{reg}}}^\infty - \ireg{0}{\infty}\od x\, fg'.
\end{equation}
By contrast, changes of variables and Fubini's theorem do not hold for regularised integrals as defined here. Fubini's theorem allows one to change the order of integration.
For regularised integrals, however, changing the integration order can alter the result.
Similarly, a naive change of variables, such as the substitution $x=\lambda y$, may not preserve the regularised value: $\ireg{0}{1}\od x/x = 0$, whereas $\ireg{0}{1/\lambda}\od y/y = -\log\lambda$.
Consequently, when extending Eq.~\eqref{eq:ireg-example} to the case of several variables, the integration order and the parametrisation must be fixed before regularisation begins, and every term must be integrated using the same choices.

Consider now a parametric integral of several variables
\begin{equation}
  \label{eq:Cheng-Wu-ordered}
  I(\varepsilon) =\int_{(0,\infty)^n} \hspace*{-15pt} \od x_1\cdots\od x_n\,f(x)\, ,
  \quad
  f(x)= \prod_k P_k(x)^{m_k+a_k\varepsilon}
  \, ,
\end{equation}
where $m_k\in\mathbb{Z}$, $a_k\in\mathbb{R}$, and the $P_k$ are polynomials with non-negative coefficients (unless $m_k>0$ and $a_k=0$).
This form covers Euclidean parametric Feynman integrals (after applying the Cheng--Wu theorem).
For the fixed integration order, with $x_1$ integrated first and $x_n$ last, regularisation then proceeds in reverse order.

We identify singular regions by analysing the integrand under scalings, which are related to weighted blow-ups.
At the step where $x_j$ is regularised, we consider $\rho=(\rho_1,\ldots,\rho_j)$ with $\rho_j\ne0$ and transform $x_i\mapsto t^{\rho_i}x_i$ for $i\leq j$.
We obtain
\begin{equation}
  f(x)\od x\,\mapsto
  t^{\sum_{i \le j}\rho_i}\, f\bigl(t^{\rho_1}x_1,\ldots, t^{\rho_j}x_j, x_{j+1},\ldots, x_n\bigr)\od x \, .
\end{equation}
The scaled integrand takes the form
\begin{equation}
  t^{\sigma(\rho)}g_\rho(x,t)\, ,
  \qquad
  g_\rho(x,0)\neq0\, ,
\end{equation}
where $\sigma(\rho)$ is the net scaling degree of the integration form,
\begin{equation}
  \sigma(\rho) =\sum_{i\le j}\rho_i+\sum_k(m_k+a_k\varepsilon)\,\ldeg_\rho P_k
  = \sigma_0(\rho) + \sigma_1(\rho) \varepsilon \, ,
\end{equation}
with $\ldeg_\rho P =\min_r\sum_{i\le j}\rho_i r_i$, where the minimum is over all multi-exponents $r=(r_1,\ldots,r_n)$ in $P$.

A direction is divergent if $\sigma_0(\rho) \le 0 $. It is logarithmic for
$\sigma_0(\rho)=0$ and power-like of order $-\sigma_0(\rho)$ for $\sigma_0(\rho)<0$.
The corresponding boundary integrand is

\begin{equation}
  f_\rho(x_1,.., x_{j-1},x_{j+1},.., x_n) =
  \lim_{t\to0,{\mathrm{reg}}}t^{\sigma_0(\rho)}g_\rho(x,t)\big|_{x_j=1}\,.
  \label{eq:boundary}
\end{equation}
The one-variable result applied in $x_j$ shows that each divergent direction $\rho$ contributes a pole $|\rho_j|/(\sigma_1(\rho)\varepsilon)$ multiplying the integral of its boundary integrand.
The numerator $|\rho_j|$ reflects that $\rho$ is a direction and the pole factor must be invariant under scaling $\rho$.

To illustrate the recursive procedure of the regularisation, consider the outermost variable $x_n$ and define the inner integral
\begin{equation}
  I_{n-1}(x_n,\varepsilon) = \int_{(0,\infty)^{n-1}} \hspace*{-10pt} \od x_1\cdots\od x_{n-1}\,f(x) \, .
\end{equation}
Its behaviour at $x_n=0$ and $x_n=\infty$ receives contributions from all divergent directions with $\rho_n\ne0$.
Each such contribution has a scaling $\sigma_1(\rho)$ and a coefficient given by an integral over $x_1,\ldots,x_{n-1}$.
Applying the regularisation in $x_n$ gives
\begin{equation}
  I(\varepsilon) = I_{\mathrm{reg}}(\varepsilon)
  + \sum_{\rho\in D_n}\frac{|\rho_n|}{\sigma_1(\rho)\varepsilon} \int_{(0,\infty)^{n-1}} \hspace*{-10pt} \od x_1\cdots\od x_{n-1}\,f_\rho(x)
  \, ,
  \label{eq:method1}
\end{equation}
where $D_n$ denotes the set of divergent directions at step $n$.
The procedure is iterated on the regularised remainder integrals.
The singularity structure of the remainder integrals can be inferred from the singularities of $f$.
The iteration yields a sum of regularised integrals, each multiplied by explicit pole factors $1/\prod_i(\sigma_1(\rho_i)\varepsilon)$.
These integrals are expanded in $\varepsilon$ and evaluated in terms of hyperlogarithms (if possible) with the expansion depth determined by the total pole order of the prefactor.

\begin{figure}[t]
  \centering
  \begin{tikzpicture}[scale=1.25, >=Stealth]

    \draw[->,gray] (0,0) -- (2.8,0)
      node[right,black] {$r_1$};
    \draw[->,gray] (0,0) -- (0,2.8)
      node[left,black] {$r_2$};

    \foreach \i in {1,2} {
      \draw[gray] (\i,-0.05) -- (\i,0.05)
        node[below=3pt,black] {\small \i};
      \draw[gray] (-0.05,\i) -- (0.05,\i)
        node[left=3pt,black] {\small \i};
    }

    \fill[blue!6]
      (0,1) -- (0,2) -- (1,2) -- (2,1)
      -- (2,0) -- (1,0) -- cycle;

    \draw[thick,gray,line join=round]
      (0,2) -- (1,2)
      -- (2,1) -- (2,0);

    \draw[very thick,red!75!black,line join=round]
      (0,2) -- (0,1)
      -- (1,0) -- (2,0);

    \foreach \p in {(0,1),(0,2),(1,2),(2,1),(2,0),(1,0),(1,1)}
      \fill[black] \p circle (1.5pt);

    \draw[->,red!75!black,thick] (0,1.5)   -- (0.45,1.5);
    \draw[->,red!75!black,thick] (1.5,0)   -- (1.5,0.45);
    \draw[->,red!75!black,thick] (0.5,0.5) -- (0.82,0.82);

    \node[red!75!black,left,align=center,inner sep=2pt] at (0,1.5)
      {$\mathcal F_1$\\[-1pt] \small$(1,0)$};
    \node[red!75!black,below,align=center,inner sep=2pt] at (1.5,0)
      {$\mathcal F_2$\\[-1pt] \small$(0,1)$};
    \node[red!75!black,below left,align=center,inner sep=1pt] at (0.67,0.57)
      {$\mathcal F_3$\\[-1pt] \small$(1,1)$};

  \end{tikzpicture}
  \caption{Newton polytope of
    $(x_1+x_2)(1+x_1)(1+x_2)$.
    The coordinates $r_1$ and $r_2$ are the exponents of the monomials.
    The three red facets are singular, with their inward normals indicating the corresponding divergent scaling directions.
    The grey facets correspond to convergent directions.}
  \label{fig:singular-facets}
\end{figure}
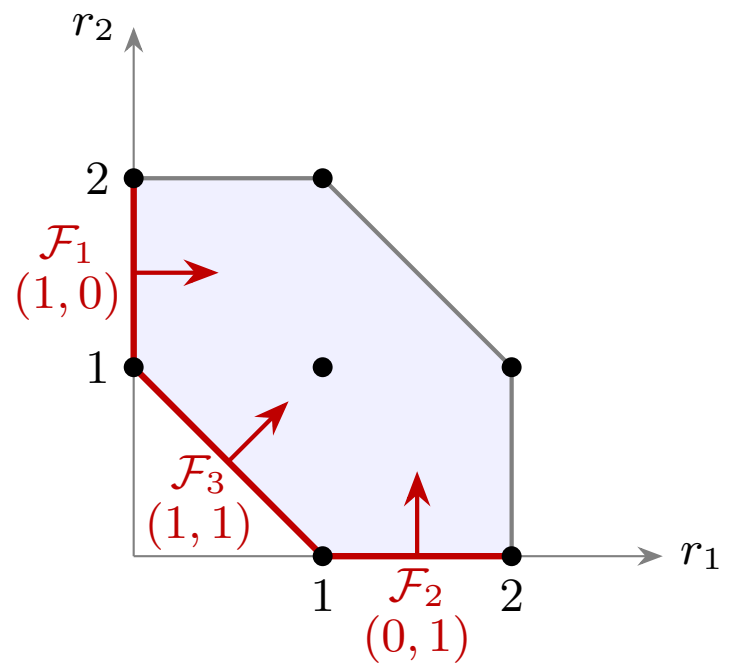
The identification of singular directions of $I(\varepsilon)$ in Eq.~\eqref{eq:Cheng-Wu-ordered} by scaling is, in fact, entirely combinatorial and follows from the facets of the Newton polytope of the integrand, as illustrated in Fig.~\ref{fig:singular-facets} for $f(x)=x_1^{-1+a_1\varepsilon}x_2^{-1+a_2\varepsilon}(x_1+x_2)^{-1+a_3\varepsilon}/((1+x_1)(1+x_2))$.
A facet is singular if its inward normal defines a divergent scaling direction.
If a singular direction does not correspond to a facet of the Newton polytope, its contribution to the regularisation can be omitted, as it amounts to an unnecessary
over-regularisation.
This restriction allows us to reduce the infinite number of divergent directions to a finite number of singular facets.

In situations with a highly nested singularity structure, the iterative algorithm
(the fibration method) becomes slow because it produces many pole factors for the same integrand.
A more advanced approach (the facet method) provides the pole factors directly. For this method, we augment each divergent normal $\rho$ of the Newton polytope with its associated $\varepsilon$-scaling
$\sigma_1(\rho)$.
The facets are then processed in reverse integration order. At each step, the extended normals are reduced to account for variables already integrated out, and only admissible directions with matching scaling contribute.
Each contributes a boundary integral and a pole factor $1/(\sigma_1(\rho)\varepsilon)$, where $\rho$ is a normal of a singular facet of the Newton polytope of $f$.
Enumerating all admissible sets of facets directly yields the regularised representation of $I(\varepsilon)$.
Details are given in Ref.~\cite{Schnetz:2026}.

As a first example, we consider the planar $\ell$-loop ladder-box integrals in $D=4-2\varepsilon$, shown in Fig.~\ref{fig:ladder}.
We take all external momenta $p_1,\dots,p_4$ to be incoming and lightlike.
Abbreviating $p_{ij}=p_i+p_j$, we have the Mandelstam variables $s=p_{12}^2$ and $t=p_{23}^2$, which are negative in the Euclidean region. By momentum conservation, $u=-s-t$. Using the momentum routing of Fig.~\ref{fig:ladder}, we define the ladder integrals in Euclidean momenta,
\begin{widetext}
\begin{equation}
  \mathcal{B}^{(\ell)}(x) = \text{e}^{\ell\gamma_\text{E}\varepsilon}\,(-s)^\ell\,(-t)^{1+\ell\varepsilon}\int\Bigg(\prod_{j=1}^\ell\frac{\od^Dk_j}{\pi^{D/2}}\Bigg)\,
  \frac{1}{(k_1-p_1)^2\,(k_\ell+p_4)^2\,\prod_{j=1}^\ell k_j^2\,(k_j-p_{12})^2\,\prod_{j=1}^{\ell-1}(k_j-k_{j+1})^2}\, .
  \label{eq:ladder-mom}
\end{equation}
\end{widetext}
We use the fact that $x=t/s$ is the only dimensionless ratio
and expand the ladder boxes as
\begin{equation}
  \mathcal{B}^{(\ell)}(x) = \sum_{k=-2\ell}^\infty\, c^{(\ell)}_k(x)\,\varepsilon^k\, .
  \label{eq:ladder-exp}
\end{equation}

We have computed the coefficients $c_k^{(\ell)}$ up to five loops, obtaining the Laurent expansions to $\varepsilon^8$, $\varepsilon^6$, $\varepsilon^4$, $\varepsilon^2$ and $\varepsilon^0$ for $\ell=1,\ldots,5$, respectively.
For the chosen kinematics, the coefficients can be expressed in terms of harmonic polylogarithms (HPLs)~\cite{Remiddi:1999ew}, $H_w\equiv H_w(x)$, with letters $0$ and $-1$, where $H_0(x)=\log x$ and $H_{-1}(x)=\log(1+x)$.
We define multiple zeta values
\begin{equation}
  \zeta_{n_1,\ldots,n_r}=\sum_{0<k_1<\ldots<k_r}\frac{1}{k_1^{n_1}\cdots k_r^{n_r}}\, ,
\end{equation}
where the value $\zeta_{3,5}$ appears for the first time in $c_4^{(2)}$.

Throughout, we use the facet-based regularised representation of $\mathcal{B}^{(\ell)}$, which can be computed in seconds using \texttt{Maple} (see Table~\ref{tab:ladder-timings}).
The subsequent iterated integrations yielding HPLs are performed with an implementation of the algorithms of Ref.~\cite{Panzer:2014caa}, built on the Rust library \texttt{Symbolica}~\cite{ruijl_symbolica_2025}.
This step is computationally efficient: It takes only a minute to obtain $\mathcal{B}^{(4)}$ to order $\varepsilon^0$.
\begin{figure}[tbp]
  \centering
  \begin{tikzpicture}[line width=0.9pt,scale=1.1,
      lab/.style={font=\scriptsize,inner sep=1pt},
      ext/.style={postaction={decorate},
        decoration={markings,mark=at position 0.6 with
          {\arrow{Stealth[length=5pt]}}}}]

    \draw (0,1.6) -- node[lab,above] {$k_1$} (1.4,1.6);
    \draw (0,0) -- node[lab,below] {$k_1-p_{12}$} (1.4,0);
    \draw[dotted] (1.4,1.6) -- (2.4,1.6);
    \draw[dotted] (1.4,0) -- (2.4,0);
    \draw (2.4,1.6) -- node[lab,above] {$k_{\ell-1}$} (4.1,1.6);
    \draw (2.4,0) -- node[lab,below] {$k_{\ell-1}-p_{12}$} (4.1,0);
    \draw (4.1,1.6) -- node[lab,above] {$k_\ell$} (5.4,1.6);
    \draw (4.1,0) -- node[lab,below] {$k_\ell-p_{12}$} (5.4,0);

    \draw (0,0) -- node[lab,right] {$k_1-p_1$} (0,1.6);
    \draw (1.4,0) -- (1.4,1.6);
    \draw (2.4,0) -- node[lab,right] {$k_{\ell-2}-k_{\ell-1}$} (2.4,1.6);
    \draw (4.1,0) -- node[lab,right] {$k_{\ell-1}-k_\ell$} (4.1,1.6);
    \draw (5.4,0) -- node[lab,right] {$k_\ell+p_4$} (5.4,1.6);

    \draw[ext] ($(0,1.6)+(-0.6,0.6)$)
      node[above left=-2pt] {$p_1$} -- (0,1.6);
    \draw[ext] ($(0,0)+(-0.6,-0.6)$)
      node[below left=-2pt] {$p_2$} -- (0,0);
    \draw[ext] ($(5.4,1.6)+(0.6,0.6)$)
      node[above right=-2pt] {$p_4$} -- (5.4,1.6);
    \draw[ext] ($(5.4,0)+(0.6,-0.6)$)
      node[below right=-2pt] {$p_3$} -- (5.4,0);

    \foreach \xx in {0,1.4,2.4,4.1,5.4} {
      \fill (\xx,0) circle (1.6pt);
      \fill (\xx,1.6) circle (1.6pt);
    }

  \end{tikzpicture}
  \caption{The $\ell$-loop ladder box with external and loop momenta assigned, using the shorthand notation $p_{12}=p_1+p_2$.}
  \label{fig:ladder}
\end{figure}
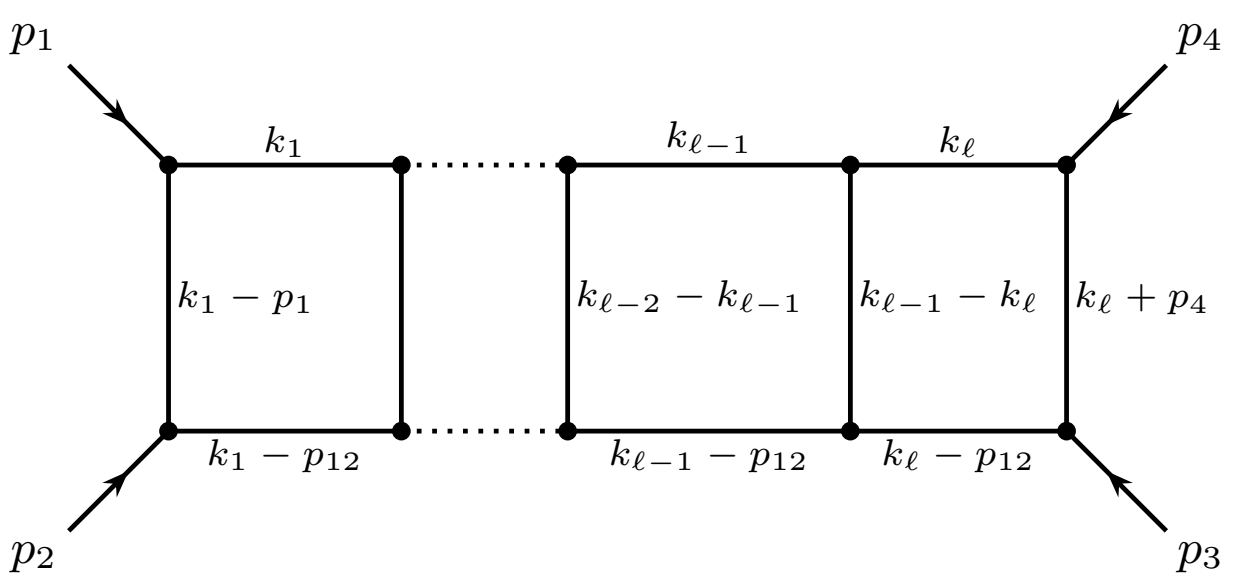
\begin{table}[tbp]
  \centering
  \footnotesize
  \setlength{\tabcolsep}{2pt}
  \begin{tabular}{@{}l r r r r r r r@{}}
    \toprule
    $\ell$        & 1      & 2      & 3      & 4      & 5      & 6     & 7                    \\
    \midrule
    terms         & 5      & 43     & 293    & 1762   & 9803   & 51738 & 262884               \\
    wall time [s] & $<\!\!1$ & $<\!\!1$ & $<\!\!1$ & $3.5$ & $24$ & $160$ & $1.2{\times}10^{3}$ \\
    \bottomrule
  \end{tabular}
  \caption{Facet regularisation of the $\ell$-loop ladder box:
    number of terms of the regularised list after canonicalisation of the
    integration variables, and wall time for generating the list
    (\texttt{Maple}, one thread). The timing refers to the facet method using the fact that the singular facets are known to all orders.
  }
  \label{tab:ladder-timings}
\end{table}

We have validated our results against several known calculations.
At one loop, we reproduce the finite expansion of the massless on-shell box integral in Ref.~\cite{Duplancic:2000sk} and the expansion of its hypergeometric representation in Ref.~\cite{Bern:1993kr} up to $\varepsilon^8$.
At two and three loops, our results agree with the planar double-box and triple-box integrals of Refs.~\cite{Smirnov:1999gc,Smirnov:2003vi}, respectively, up to $\varepsilon^0$.
This comparison requires converting to the Minkowski normalisation and kinematic conventions used in those references.
For the triple box, we additionally perform a change of kinematic variable and map the HPLs accordingly.
Finally, we perform a numerical evaluation by sector decomposition with \texttt{pySecDec}~\cite{Heinrich:2023til}, confirming all orders of the four-loop ladder up to $\varepsilon^0$, with relative numerical uncertainties below $1\%$ up to $\varepsilon^{-4}$ and around $5\%$ at $\varepsilon^{-3}$ to order $\varepsilon^{0}$.
Additionally, we compared our result with the partial results for its finite, $x$-dependent part presented in Ref.~\cite{Bargiela:2024hgt}, which were obtained with the standard approach based on IBP reduction and differential equations.
All $x$-dependent coefficients agree except for the coefficient multiplying $\log(-x)$ at order $\varepsilon^0$.
Its imaginary part, as listed in Ref.~\cite{Bargiela:2024hgt}, is incompatible with the cancellation required for the full result to be real in Euclidean kinematics.
Recently, canonical differential equations for $\mathcal{B}^{(4)}$ and its subtopologies have been derived using intersection theory~\cite{Brunello:2026vij}.
Our results for the Laurent expansion of $\mathcal{B}^{(\ell)}$ in Eq.~\eqref{eq:ladder-exp} to five loops are listed in the supplemental material.

For loop orders $\ell=6,7,8$, we have obtained $\mathcal{B}^{(\ell)}$ to orders $\varepsilon^{-6}$, $\varepsilon^{-10}$ and $\varepsilon^{-13}$, respectively.
These results allow us to conjecture closed-form all-loop-order expressions for the four leading Laurent coefficients, $c_{-2\ell}^{(\ell)}$ to $c_{-2\ell+3}^{(\ell)}$.
These coefficients take particularly simple forms. In the normalisation of Eq.~\eqref{eq:ladder-mom}, we find for the two leading coefficients
\begin{equation}
  \label{eq:planar-leading-poles}
  c_{-2\ell}^{(\ell)} = \frac{4^\ell}{(\ell!)^2} \,,
  \quad
  c_{-2\ell+1}^{(\ell)} = \frac{\ell(2\ell)!}{(\ell!)^4}\,H_0 \, .
\end{equation}
The observed patterns can be resummed into a generating function.
To this end, we introduce the formal loop-counting parameter $g$ and $v = 2\sqrt{g r_\Gamma }/\varepsilon$, with $r_\Gamma = \mathrm{e}^{\gamma_\mathrm{E}\varepsilon} \Gamma(1+\varepsilon)\Gamma(1-\varepsilon)^2/\Gamma(1-2\varepsilon)$.
We denote by $I_n\equiv I_n(v)$ the modified Bessel functions of the first kind and write $J_n=I_n(2v)$.
The generating function
\begin{equation}
  \mathcal{G}(g,\varepsilon;x) =\sum_{\ell\geq0}g^\ell\, \mathcal{B}^{(\ell)}(x)\,,\qquad \mathcal{B}^{(0)}\equiv1\,,
\end{equation}
then takes the form
\begin{align}
  \mathcal{G}(g,\varepsilon;x) \!=
  & J_0+\varepsilon
  v\,I_0I_1H_0+\frac{\varepsilon^2\zeta_2}{2}\Big\{\! v^2J_2-2\,\partial_v\!\big(v^2I_0I_1\big)\! \Big\}\nonumber\\
  &\hspace*{-18mm} -\frac{\varepsilon^3}{2}\Big\{v^2I_0^2\big(H_{-1,0,0}+3\zeta_2H_{-1}\big) -\zeta_2v^2I_1\,(vI_0+I_1)H_0\nonumber\\
  &\hspace*{-18mm} +\frac{\zeta_3}{3}\Big[v^3J_3-3v^2J_2+6v\,I_1\,(vI_1+I_0)\Big]\Big\}+\mathcal O(\varepsilon^4) \, ,
  \label{eq:GF}
\end{align}
where higher orders in $\varepsilon$ are to be understood at fixed $v$.

\begin{figure}[tbp]
  \centering
  \begin{tikzpicture}[line width=0.9pt,scale=1.1,
      lab/.style={font=\scriptsize,inner sep=1pt},
      ext/.style={postaction={decorate},
        decoration={markings,mark=at position 0.6 with
          {\arrow{Stealth[length=5pt]}}}}]

    \draw (0,1.6) -- node[lab,above] {$k_1$} (1.4,1.6);
    \draw (0,0) -- node[lab,below] {$k_1-p_{12}$} (1.4,0);
    \draw[dotted] (1.4,1.6) -- (2.4,1.6);
    \draw[dotted] (1.4,0) -- (2.4,0);
    \draw (2.4,1.6) -- node[lab,above] {$k_{\ell-1}$} (3.8,1.6);
    \draw (2.4,0) -- node[lab,below] {$k_{\ell-1}-p_{12}$} (3.8,0);
    \draw (3.8,1.6) -- node[lab,above] {$k_{\ell-1}-k_\ell$} (5.1,1.6);
    \draw (3.8,0) -- node[lab,below] {$k_\ell+p_3$} (5.1,0);

    \draw (0,0) -- node[lab,right] {$k_1-p_1$} (0,1.6);
    \draw (1.4,0) -- (1.4,1.6);
    \draw (2.4,0) -- node[lab,right] {$k_{\ell-2}-k_{\ell-1}$} (2.4,1.6);

    \draw (3.8,0) -- ($(3.8,0)!0.45!(5.1,1.6)$);
    \draw ($(3.8,0)!0.55!(5.1,1.6)$) -- node[lab,below right=1pt] {$k_{\ell-1}-k_\ell+p_4$} (5.1,1.6);
    \draw (3.8,1.6) -- node[lab,pos=0.75,above right=1pt] {$k_\ell$} (5.1,0);

    \draw[ext] ($(0,1.6)+(-0.6,0.6)$)
      node[above left=-2pt] {$p_1$} -- (0,1.6);
    \draw[ext] ($(0,0)+(-0.6,-0.6)$)
      node[below left=-2pt] {$p_2$} -- (0,0);
    \draw[ext] ($(5.1,1.6)+(0.6,0.6)$)
      node[above right=-2pt] {$p_4$} -- (5.1,1.6);
    \draw[ext] ($(5.1,0)+(0.6,-0.6)$)
      node[below right=-2pt] {$p_3$} -- (5.1,0);

    \foreach \xx in {0,1.4,2.4,3.8,5.1} {
      \fill (\xx,0) circle (1.6pt);
      \fill (\xx,1.6) circle (1.6pt);
    }

  \end{tikzpicture}
  \caption{The $\ell$-loop non-planar box, using the same conventions as in Fig.~\ref{fig:ladder}.}
  \label{fig:np-box}
\end{figure}
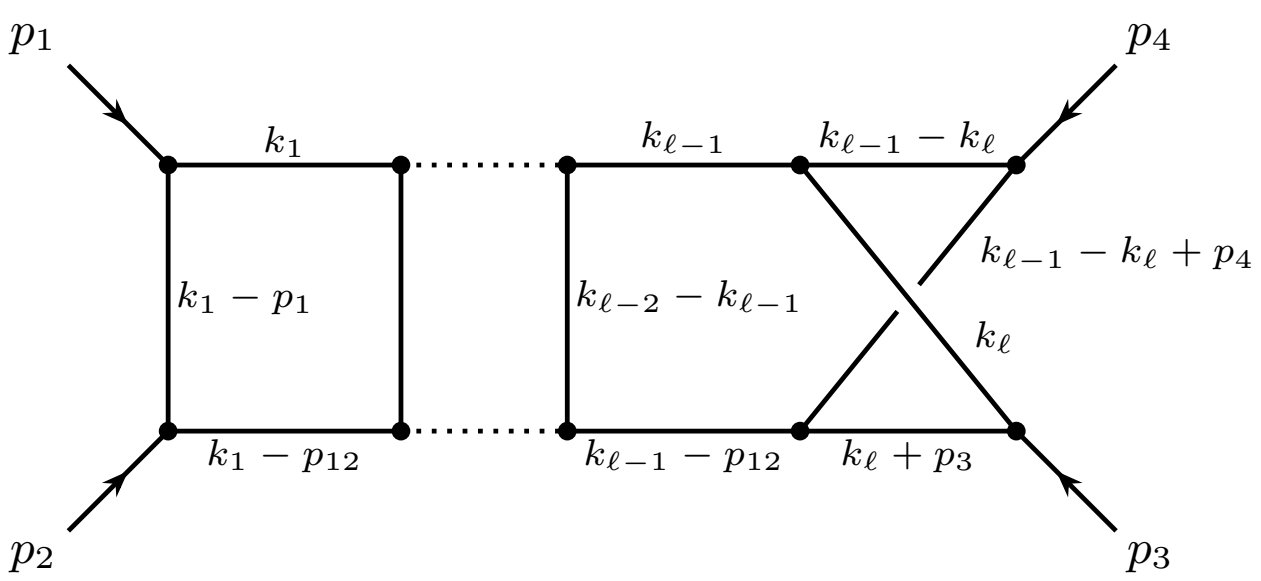
As a second example, we consider the non-planar $\ell$-loop box integrals, illustrated in Fig.~\ref{fig:np-box}.
Following the conventions of the ladder boxes in Eq.~\eqref{eq:ladder-mom}, we use the same propagator ordering and keep all external momenta $p_1,\dots,p_4$ incoming and lightlike.
We define the expansion
\begin{align}
  \BNP^{(\ell)}(x,y)
  = \sum_{k= - 2\ell}^\infty\, d^{(\ell)}_k(x,y) \, \varepsilon^k\, ,
  \label{eq:np-box}
\end{align}
where $x=t/s$, $y=u/s$ with $s=p_{12}^2$, $t=p_{23}^2$ and $u=p_{13}^2$ (the assignments of $t$ and $u$ are interchanged relative to Ref.~\cite{Tausk:1999vh}).
We choose the $s$-channel normalisation, $-s=1$, which preserves the symmetry under $p_3\leftrightarrow p_4$, implying $\BNP(x,y)=\BNP(y,x)$.
Following Ref.~\cite{Tausk:1999vh}, we initially treat $x$ and $y$ as independent positive variables, corresponding to the Euclidean region $s,t,u<0$.
Momentum conservation, $s+t+u=0$, or equivalently, $1+x+y=0$, is imposed only at the end by analytic continuation.

We have obtained the coefficients $d_k^{(\ell)}$ in Eq.~\eqref{eq:np-box} to orders $\varepsilon^4$, $\varepsilon^2$, and $\varepsilon^0$ for $\ell=2,3,4$, respectively, and verified the exchange symmetry $x\leftrightarrow y$ at all available orders.
The analytic results are listed in the supplemental material. They are expressed in terms of multiple polylogarithms~\cite{Goncharov:1998kja}, which generalise the HPLs used above.
They contain letters $\{0,-1\}$ in $y$ and $\{0,-1,-y,-1-y\}$ in $x$, which correspond to the singular loci $x=0$, $1+x=0$, $y=0$, $1+y=0$, $x+y=0$ and $1+x+y=0$.

The two-loop non-planar double box $\BNP^{(2)}$ was computed up to
$\varepsilon^0$ in Ref.~\cite{Tausk:1999vh}, and we find full agreement.
We analytically continued the three-loop result $\BNP^{(3)}$ to the physical point $(x,y)=(\tfrac13,-\tfrac43)$.
There, it agrees with the IBP-reduced integral to 36 significant digits at every order in $\varepsilon$.
The reduction to the master integrals of Ref.~\cite{Henn:2020lye} was performed with \textsc{Kira}~\cite{Lange:2025fba}.

We have also performed a facet-based computation of the leading poles of $\BNP^{(5)}$ and $\BNP^{(6)}$. All leading poles are rational functions of $x$ and $y$.
In the normalisation of Eq.~\eqref{eq:np-box}, we obtain
\begin{equation}
  \begin{aligned}
    xy\,d^{(2)}_{-4} &= \frac{5+x+y}{2}\,,\\
    xy\,d^{(3)}_{-6} &= \frac{22+33\,(x+y)}{45}\,,\\
    xy\,d^{(4)}_{-8} &= \frac{93+599\,(x+y)}{2016}\,, \\
    xy\,d^{(5)}_{-10} &= \frac{193+4465\,(x+y)}{75600}\,,\\
    xy\,d^{(6)}_{-12} &= \frac{793+62021\,(x+y)}{8553600}\,.
  \end{aligned}
  \label{eq:np-leading-poles}
\end{equation}
Using $x+y=-1$, the leading pole of $\BNP^{(\ell)}$ multiplied by $xy$ is independent of the kinematics.
The constant terms in Eq.~\eqref{eq:np-leading-poles} can be written as
\begin{equation}
  xy\,d^{(\ell)}_{-2\ell}\big|_{x=y=0} = 2(\ell+1)\left[\frac{4^\ell}{(2\ell)!}-\frac{1}{(\ell!)^2}\right],
  \label{eq:np-leading-poles-const}
\end{equation}
which we conjecture to hold at all loop orders.
For the coefficient of $x+y$, and hence for the on-shell leading pole, we have not been able to determine a closed form.

The regularisation method is not restricted to massless on-shell $2\to2$ scattering.
It can also be used for integrals with off-shell external legs, $p_i^2\neq0$, providing an alternative approach to recent calculations of three-loop massless four-point integrals with one off-shell leg~\cite{Gehrmann:2026mln}.
Extensions to $2\to3$ and $2\to4$ scattering processes are also within the method's scope.
These applications involve additional independent kinematic invariants and, consequently, larger alphabets of hyperlogarithmic letters.
At five points, the relevant letters include those of the planar pentagon alphabet~\cite{Gehrmann:2015bfy}.
Momentum-twistor variables play an important role in rationalising Gram-determinant square roots and establishing linear reducibility with symbolic kinematics~\cite{Henn:2024sixpoint}.

We have tested the method on additional five- and six-point integrals.
At five points, we computed the massless one-loop pentagon to  $\varepsilon^2$, the scalar and dual-conformal two-loop pentaboxes to $\varepsilon^{-2}$, and their six- and seven-propagator subtopologies to $\varepsilon^0$.
In these calculations, we have used the Lee--Pomeransky representation~\cite{Lee:2013hzt} when necessary to restore linear reducibility.
All obtained results agree with the available literature~\cite{Bern:1993kr,Papadopoulos:2015jft}.
At six points, we find excellent numerical agreement with the known result for the finite one-loop hexagon in six dimensions~\cite{Dixon:2011ng}.
We also applied the method to two planar two-loop six-point sectors in general kinematics, the hexagon-bubble and the pentagon-triangle, obtaining their leading two Laurent coefficients in agreement with results from \texttt{pySecDec}~\cite{Heinrich:2023til}.
A comprehensive study of these higher-point examples, in particular the six-point two-loop sectors, is left for future work.

Another promising direction for future work is the extension of this framework to Feynman integrals with massive internal lines. This poses additional challenges in general kinematics where individual terms may fail to be linearly reducible, so that developing strategies to overcome these limitations would
further broaden the scope and applicability of the method.

Parametric integration combined with our regularisation procedure provides an efficient approach to direct computation of Feynman integrals.
The pole structure is exposed explicitly through boundary contributions, with each extraction reducing the number of remaining integrations.
The finite parts are then obtained through regularised iterated integrations, exploiting linear reducibility and hyperlogarithmic representations.
All steps are algorithmic and implemented in computer algebra systems, requiring comparatively modest computational resources in contrast to the conventional approach based on IBP reduction to master integrals and the solution of differential equations.

\vspace{2mm}
\begin{acknowledgments}
\emph{Acknowledgements:}
We thank Bakar Chargeishvili for discussions.
We acknowledge the use of Claude (Anthropic) and ChatGPT (OpenAI) for code development and for writing assistance.

This work has been supported by the Deutsche Forschungsgemeinschaft (DFG) under the grants EXC 2121 {\it Quantum Universe} 390833306, HPC cluster {\it Hummel-2} 498394658, and SCHN 1240/3-1, by the University of Debrecen Program for Scientific Publication, by the National Research, Development and Innovation Office (NKFIH, Hungary) research grant 2026-4.1.1-MISSZI\'OK-2026-00005, and by the European Research Council (ERC) under the ERC Advanced Grant 101095857 {\it Conformal-EIC} and the ERC Synergy Grant 101167287 {\it MaScAmp}.

Ancillary files with our results can be obtained from the preprint server \url{http://arXiv.org} and are also available from the authors upon request.

\end{acknowledgments}


\providecommand{\href}[2]{#2}\begingroup\raggedright\endgroup

\newpage
\cleardoublepage
\appendix
\renewcommand{\theequation}{A.\arabic{equation}}
\setcounter{equation}{0}
\renewcommand{\thefigure}{A.\arabic{figure}}
\setcounter{figure}{0}
\renewcommand{\thetable}{A.\arabic{table}}
\setcounter{table}{0}

\begin{widetext}
\section{Supplemental material}
\subsection{Ladder boxes}
\noindent We expand the results as
\begin{equation}
  \mathcal{B}^{(\ell)}(x) = \sum_{k=-2\ell}^\infty c_{k}^{(\ell)}(x)\,\varepsilon^{k} \,,
  \label{eq:ladder-exp-A}
\end{equation}
and write $H_w\equiv H_w(x)$ for the harmonic polylogarithms~\cite{Remiddi:1999ew}, defined recursively by $H_{a,w}(x)=\int_0^x\od t\,f_a(t)\,H_{w}(t)$ with $f_0(t)=1/t$, $f_{-1}(t)=1/(1+t)$ and $H_{\textbf{e}}(x)=1$, where \textbf{e} denotes the empty word. For words of $n$ zeros only, $H_{0,\dots,0}(x)=\frac{1}{n!}\log^n x$.
We obtain $\mathcal{B}^{(\ell)}$ through $\varepsilon^{8}$, $\varepsilon^{6}$, $\varepsilon^{4}$, $\varepsilon^{2}$ and $\varepsilon^{0}$ for $\ell=1,\dots,5$, respectively.

\subsubsection{Leading poles}

\noindent The four leading coefficients follow simple patterns. With
\begin{equation}
  a_\ell \equiv \frac{(2\ell)!}{(\ell!)^4} = \frac{1}{(\ell!)^2}\binom{2\ell}{\ell}\,,\qquad a_0=1\,,
\end{equation}
we find for $\ell=1,\dots,8$
\begin{align}
  c_{-2\ell}^{(\ell)}\quad &= \frac{4^\ell}{(\ell!)^2} \,, \label{eq:lp0}\\
  c_{-2\ell+1}^{(\ell)} &= \ell\,a_\ell\,H_0 \,, \label{eq:lp1}\\
  c_{-2\ell+2}^{(\ell)} &= \Big[\frac{\ell(\ell-2)}{2}\,c_{-2\ell}^{(\ell)}-(2\ell+1)\,\ell\,a_\ell\Big]\zeta_2 \,, \label{eq:lp2}\\
  c^{(\ell)}_{-2\ell+3} &= -a_{\ell-1}\Big[2H_{-1,0,0}+6\zeta_2H_{-1}-\frac{2\ell^2-2\ell-1}{\ell}\,\zeta_2H_0+\frac{2(2\ell^2-1)}{\ell^2}\,\zeta_3\Big]-\frac{\ell(\ell^2-6\ell+19)}{6}\,c^{(\ell)}_{-2\ell}\,\zeta_3 \, . \label{eq:lp3}
\end{align}

\noindent We conjecture the patterns to hold to all loop orders. We note that the coefficient $c^{(\ell)}_{-2\ell+1}$ and the first bracket of $c^{(\ell)}_{-2\ell+3}$ involve the same sequence $a_\ell$, shifted by one loop order, and the $\zeta_3$ term of $c^{(\ell)}_{-2\ell+3}$ has the same structure as the $\zeta_2$ term of $c^{(\ell)}_{-2\ell+2}$.

\subsubsection{Complete results}

\noindent For $\ell=1,\dots,5$ we display the first orders in $\varepsilon$ below, with the multiple zeta value $\zeta_{3,5}=\sum_{0<k_1<k_2}k_1^{-3}k_2^{-5}$. To keep the material at a reasonable length, we do not show all calculated orders here. All orders obtained can be found in the ancillary files.
\allowdisplaybreaks
\begin{align}
  c^{(1)}_{-2} &= 4\,, \notag\\
  c^{(1)}_{-1} &= 2H_{0}\,, \notag\\
  c^{(1)}_{0} &= - \tfrac{4}{3}\pi^2\,, \notag\\
  c^{(1)}_{1} &= - 2H_{-1,0,0} - \tfrac{1}{6}\pi^2H_{0} - \pi^2H_{-1} - \tfrac{34}{3}\zeta_3\,, \notag\\
  c^{(1)}_{2} &= - 2H_{-1,0,0,0} - 2H_{-1,-1,0,0} - \pi^2H_{-1,-1} - \tfrac{14}{3}\zeta_3H_{0} - 2\zeta_3H_{-1} - \tfrac{41}{360}\pi^4\,, \notag\\
  c^{(1)}_{3} &= - 2H_{-1,0,0,0,0} - 2H_{-1,-1,0,0,0} - 2H_{-1,-1,-1,0,0} + \tfrac{1}{6}\pi^2H_{-1,0,0} - \pi^2H_{-1,-1,-1} - 2\zeta_3H_{-1,-1} - \tfrac{47}{720}\pi^4H_{0} \notag\\
  &\quad + \tfrac{1}{60}\pi^4H_{-1} + \tfrac{59}{18}\zeta_3\pi^2 - \tfrac{134}{5}\zeta_5\,, \notag\\
  c^{(1)}_{4} &= - 2H_{-1,0,0,0,0,0} - 2H_{-1,-1,0,0,0,0} - 2H_{-1,-1,-1,0,0,0} - 2H_{-1,-1,-1,-1,0,0} + \tfrac{1}{6}\pi^2H_{-1,0,0,0} \notag\\
  &\quad + \tfrac{1}{6}\pi^2H_{-1,-1,0,0} - \pi^2H_{-1,-1,-1,-1} + \tfrac{14}{3}\zeta_3H_{-1,0,0} - 2\zeta_3H_{-1,-1,-1} + \tfrac{1}{60}\pi^4H_{-1,-1} + \tfrac{7}{18}\zeta_3\pi^2H_{0} \notag\\
  &\quad + \tfrac{5}{2}\zeta_3\pi^2H_{-1} - \tfrac{62}{5}\zeta_5H_{0} - 2\zeta_5H_{-1} + \tfrac{140}{9}\zeta_3^2 + \tfrac{1}{2160}\pi^6\,, \notag\\
  c^{(1)}_{5} &= - 2H_{-1,0,0,0,0,0,0} - 2H_{-1,-1,0,0,0,0,0} - 2H_{-1,-1,-1,0,0,0,0} - 2H_{-1,-1,-1,-1,0,0,0} - 2H_{-1,-1,-1,-1,-1,0,0} \notag\\
  &\quad + \tfrac{1}{6}\pi^2H_{-1,0,0,0,0} + \tfrac{1}{6}\pi^2H_{-1,-1,0,0,0} + \tfrac{1}{6}\pi^2H_{-1,-1,-1,0,0} - \pi^2H_{-1,-1,-1,-1,-1} + \tfrac{14}{3}\zeta_3H_{-1,0,0,0} \notag\\
  &\quad + \tfrac{14}{3}\zeta_3H_{-1,-1,0,0} - 2\zeta_3H_{-1,-1,-1,-1} + \tfrac{47}{720}\pi^4H_{-1,0,0} + \tfrac{1}{60}\pi^4H_{-1,-1,-1} + \tfrac{5}{2}\zeta_3\pi^2H_{-1,-1} - 2\zeta_5H_{-1,-1} \notag\\
  &\quad + \tfrac{49}{9}\zeta_3^2H_{0} + \tfrac{14}{3}\zeta_3^2H_{-1} - \tfrac{949}{60480}\pi^6H_{0} + \tfrac{107}{3360}\pi^6H_{-1} + \tfrac{143}{432}\zeta_3\pi^4 + \tfrac{253}{30}\zeta_5\pi^2 - \tfrac{522}{7}\zeta_7\,, \notag\\
  c^{(1)}_{6} &= - 2H_{-1,0,0,0,0,0,0,0} - 2H_{-1,-1,0,0,0,0,0,0} - 2H_{-1,-1,-1,0,0,0,0,0} - 2H_{-1,-1,-1,-1,0,0,0,0} \notag\\
  &\quad - 2H_{-1,-1,-1,-1,-1,0,0,0} - 2H_{-1,-1,-1,-1,-1,-1,0,0} + \tfrac{1}{6}\pi^2H_{-1,0,0,0,0,0} + \tfrac{1}{6}\pi^2H_{-1,-1,0,0,0,0} \notag\\
  &\quad + \tfrac{1}{6}\pi^2H_{-1,-1,-1,0,0,0} + \tfrac{1}{6}\pi^2H_{-1,-1,-1,-1,0,0} - \pi^2H_{-1,-1,-1,-1,-1,-1} + \tfrac{14}{3}\zeta_3H_{-1,0,0,0,0} \notag\\
  &\quad + \tfrac{14}{3}\zeta_3H_{-1,-1,0,0,0} + \tfrac{14}{3}\zeta_3H_{-1,-1,-1,0,0} - 2\zeta_3H_{-1,-1,-1,-1,-1} + \tfrac{47}{720}\pi^4H_{-1,0,0,0} + \tfrac{47}{720}\pi^4H_{-1,-1,0,0} \notag\\
  &\quad + \tfrac{1}{60}\pi^4H_{-1,-1,-1,-1} - \tfrac{7}{18}\zeta_3\pi^2H_{-1,0,0} + \tfrac{5}{2}\zeta_3\pi^2H_{-1,-1,-1} + \tfrac{62}{5}\zeta_5H_{-1,0,0} - 2\zeta_5H_{-1,-1,-1} \notag\\
  &\quad + \tfrac{14}{3}\zeta_3^2H_{-1,-1} + \tfrac{107}{3360}\pi^6H_{-1,-1} + \tfrac{329}{2160}\zeta_3\pi^4H_{0} + \tfrac{19}{720}\zeta_3\pi^4H_{-1} + \tfrac{31}{30}\zeta_5\pi^2H_{0} + \tfrac{191}{30}\zeta_5\pi^2H_{-1} - \tfrac{254}{7}\zeta_7H_{0} \notag\\
  &\quad - 2\zeta_7H_{-1} + \tfrac{1124}{15}\zeta_3\zeta_5 - \tfrac{217}{54}\zeta_3^2\pi^2 + \tfrac{1079}{483840}\pi^8\,. \label{eq:B1}
\end{align}

\begin{align}
  c^{(2)}_{-4} &= 4\,, \notag\\
  c^{(2)}_{-3} &= 3H_{0}\,, \notag\\
  c^{(2)}_{-2} &= - \tfrac{5}{2}\pi^2\,, \notag\\
  c^{(2)}_{-1} &= - 4H_{-1,0,0} + \tfrac{1}{2}\pi^2H_{0} - 2\pi^2H_{-1} - \tfrac{65}{3}\zeta_3\,, \notag\\
  c^{(2)}_{0} &= 12H_{0,-1,0,0} - 8H_{-1,0,0,0} - 4H_{-1,-1,0,0} + 6\pi^2H_{0,-1} - \tfrac{2}{3}\pi^2H_{-1,0} - 2\pi^2H_{-1,-1} - 14\zeta_3H_{0} - 4\zeta_3H_{-1} \notag\\
  &\quad - \tfrac{29}{30}\pi^4\,, \notag\\
  c^{(2)}_{1} &= 24H_{0,-1,0,0,0} + 36H_{0,-1,-1,0,0} - 16H_{-1,0,0,0,0} + 16H_{-1,0,-1,0,0} - 8H_{-1,-1,0,0,0} - 4H_{-1,-1,-1,0,0} \notag\\
  &\quad - 2\pi^2H_{0,-1,0} + 18\pi^2H_{0,-1,-1} - \tfrac{10}{3}\pi^2H_{-1,0,0} + 8\pi^2H_{-1,0,-1} - \tfrac{2}{3}\pi^2H_{-1,-1,0} - 2\pi^2H_{-1,-1,-1} \notag\\
  &\quad + 36\zeta_3H_{0,-1} - 36\zeta_3H_{-1,0} - 4\zeta_3H_{-1,-1} - \tfrac{7}{40}\pi^4H_{0} - \tfrac{17}{18}\pi^4H_{-1} + \tfrac{73}{6}\zeta_3\pi^2 - \tfrac{383}{5}\zeta_5\,, \notag\\
  c^{(2)}_{2} &= 48H_{0,-1,0,0,0,0} + 72H_{0,-1,-1,0,0,0} + 84H_{0,-1,-1,-1,0,0} - 32H_{-1,0,0,0,0,0} + 32H_{-1,0,-1,0,0,0} \notag\\
  &\quad + 64H_{-1,0,-1,-1,0,0} - 16H_{-1,-1,0,0,0,0} + 16H_{-1,-1,0,-1,0,0} - 8H_{-1,-1,-1,0,0,0} - 4H_{-1,-1,-1,-1,0,0} \notag\\
  &\quad - 6\pi^2H_{0,-1,0,0} - 2\pi^2H_{0,-1,-1,0} + 42\pi^2H_{0,-1,-1,-1} - \tfrac{20}{3}\pi^2H_{-1,0,0,0} - \tfrac{16}{3}\pi^2H_{-1,0,-1,0} \notag\\
  &\quad + 32\pi^2H_{-1,0,-1,-1} - \tfrac{10}{3}\pi^2H_{-1,-1,0,0} + 8\pi^2H_{-1,-1,0,-1} - \tfrac{2}{3}\pi^2H_{-1,-1,-1,0} - 2\pi^2H_{-1,-1,-1,-1} \notag\\
  &\quad - 12\zeta_3H_{0,-1,0} + 84\zeta_3H_{0,-1,-1} - \tfrac{232}{3}\zeta_3H_{-1,0,0} + 64\zeta_3H_{-1,0,-1} - 36\zeta_3H_{-1,-1,0} - 4\zeta_3H_{-1,-1,-1} \notag\\
  &\quad + \tfrac{23}{30}\pi^4H_{0,-1} - \tfrac{2}{3}\pi^4H_{-1,0} - \tfrac{17}{18}\pi^4H_{-1,-1} - \tfrac{25}{3}\zeta_3\pi^2H_{0} + \tfrac{16}{3}\zeta_3\pi^2H_{-1} - \tfrac{366}{5}\zeta_5H_{0} - 16\zeta_5H_{-1} + \tfrac{275}{9}\zeta_3^2 \notag\\
  &\quad - \tfrac{2357}{15120}\pi^6\,, \notag\\
  c^{(2)}_{3} &= 96H_{0,-1,0,0,0,0,0} + 144H_{0,-1,-1,0,0,0,0} + 168H_{0,-1,-1,-1,0,0,0} + 180H_{0,-1,-1,-1,-1,0,0} \notag\\
  &\quad - 64H_{-1,0,0,0,0,0,0} + 48H_{-1,0,0,-1,-1,0,0} + 64H_{-1,0,-1,0,0,0,0} - 16H_{-1,0,-1,0,-1,0,0} \notag\\
  &\quad + 128H_{-1,0,-1,-1,0,0,0} + 160H_{-1,0,-1,-1,-1,0,0} - 32H_{-1,-1,0,0,0,0,0} + 32H_{-1,-1,0,-1,0,0,0} \notag\\
  &\quad + 64H_{-1,-1,0,-1,-1,0,0} - 16H_{-1,-1,-1,0,0,0,0} + 16H_{-1,-1,-1,0,-1,0,0} - 8H_{-1,-1,-1,-1,0,0,0} \notag\\
  &\quad - 4H_{-1,-1,-1,-1,-1,0,0} - 12\pi^2H_{0,-1,0,0,0} - 10\pi^2H_{0,-1,-1,0,0} - 2\pi^2H_{0,-1,-1,-1,0} \notag\\
  &\quad + 90\pi^2H_{0,-1,-1,-1,-1} - \tfrac{40}{3}\pi^2H_{-1,0,0,0,0} - 8\pi^2H_{-1,0,0,-1,0} + 24\pi^2H_{-1,0,0,-1,-1} - \tfrac{32}{3}\pi^2H_{-1,0,-1,0,0} \notag\\
  &\quad - 8\pi^2H_{-1,0,-1,0,-1} - \tfrac{16}{3}\pi^2H_{-1,0,-1,-1,0} + 80\pi^2H_{-1,0,-1,-1,-1} - \tfrac{20}{3}\pi^2H_{-1,-1,0,0,0} \notag\\
  &\quad - \tfrac{16}{3}\pi^2H_{-1,-1,0,-1,0} + 32\pi^2H_{-1,-1,0,-1,-1} - \tfrac{10}{3}\pi^2H_{-1,-1,-1,0,0} + 8\pi^2H_{-1,-1,-1,0,-1} \notag\\
  &\quad - \tfrac{2}{3}\pi^2H_{-1,-1,-1,-1,0} - 2\pi^2H_{-1,-1,-1,-1,-1} - 152\zeta_3H_{0,-1,0,0} - 12\zeta_3H_{0,-1,-1,0} + 180\zeta_3H_{0,-1,-1,-1} \notag\\
  &\quad - \tfrac{464}{3}\zeta_3H_{-1,0,0,0} + 48\zeta_3H_{-1,0,0,-1} + 160\zeta_3H_{-1,0,-1,-1} - \tfrac{232}{3}\zeta_3H_{-1,-1,0,0} + 64\zeta_3H_{-1,-1,0,-1} \notag\\
  &\quad - 36\zeta_3H_{-1,-1,-1,0} - 4\zeta_3H_{-1,-1,-1,-1} - \tfrac{4}{15}\pi^4H_{0,-1,0} + \tfrac{59}{30}\pi^4H_{0,-1,-1} - \tfrac{3}{2}\pi^4H_{-1,0,0} + \tfrac{92}{45}\pi^4H_{-1,0,-1} \notag\\
  &\quad - \tfrac{2}{3}\pi^4H_{-1,-1,0} - \tfrac{17}{18}\pi^4H_{-1,-1,-1} - 72\zeta_3\pi^2H_{0,-1} + \tfrac{82}{9}\zeta_3\pi^2H_{-1,0} + \tfrac{16}{3}\zeta_3\pi^2H_{-1,-1} + 168\zeta_5H_{0,-1} \notag\\
  &\quad - 252\zeta_5H_{-1,0} - 16\zeta_5H_{-1,-1} - \tfrac{64}{3}\zeta_3^2H_{0} + \tfrac{2}{3}\zeta_3^2H_{-1} - \tfrac{1541}{5040}\pi^6H_{0} - \tfrac{37}{315}\pi^6H_{-1} + \tfrac{2623}{360}\zeta_3\pi^4 + \tfrac{203}{2}\zeta_5\pi^2 \notag\\
  &\quad - \tfrac{2129}{7}\zeta_7\,, \notag\\
  c^{(2)}_{4} &= 192H_{0,-1,0,0,0,0,0,0} + 288H_{0,-1,-1,0,0,0,0,0} + 336H_{0,-1,-1,-1,0,0,0,0} + 360H_{0,-1,-1,-1,-1,0,0,0} \notag\\
  &\quad + 372H_{0,-1,-1,-1,-1,-1,0,0} - 128H_{-1,0,0,0,0,0,0,0} + 144H_{-1,0,0,0,-1,-1,0,0} - 48H_{-1,0,0,-1,0,-1,0,0} \notag\\
  &\quad + 96H_{-1,0,0,-1,-1,0,0,0} + 144H_{-1,0,0,-1,-1,-1,0,0} + 128H_{-1,0,-1,0,0,0,0,0} - 32H_{-1,0,-1,0,-1,0,0,0} \notag\\
  &\quad - 64H_{-1,0,-1,0,-1,-1,0,0} + 256H_{-1,0,-1,-1,0,0,0,0} - 16H_{-1,0,-1,-1,0,-1,0,0} + 320H_{-1,0,-1,-1,-1,0,0,0} \notag\\
  &\quad + 352H_{-1,0,-1,-1,-1,-1,0,0} - 64H_{-1,-1,0,0,0,0,0,0} + 48H_{-1,-1,0,0,-1,-1,0,0} + 64H_{-1,-1,0,-1,0,0,0,0} \notag\\
  &\quad - 16H_{-1,-1,0,-1,0,-1,0,0} + 128H_{-1,-1,0,-1,-1,0,0,0} + 160H_{-1,-1,0,-1,-1,-1,0,0} - 32H_{-1,-1,-1,0,0,0,0,0} \notag\\
  &\quad + 32H_{-1,-1,-1,0,-1,0,0,0} + 64H_{-1,-1,-1,0,-1,-1,0,0} - 16H_{-1,-1,-1,-1,0,0,0,0} + 16H_{-1,-1,-1,-1,0,-1,0,0} \notag\\
  &\quad - 8H_{-1,-1,-1,-1,-1,0,0,0} - 4H_{-1,-1,-1,-1,-1,-1,0,0} - 24\pi^2H_{0,-1,0,0,0,0} - 20\pi^2H_{0,-1,-1,0,0,0} \notag\\
  &\quad - 18\pi^2H_{0,-1,-1,-1,0,0} - 2\pi^2H_{0,-1,-1,-1,-1,0} + 186\pi^2H_{0,-1,-1,-1,-1,-1} - \tfrac{80}{3}\pi^2H_{-1,0,0,0,0,0} \notag\\
  &\quad - 24\pi^2H_{-1,0,0,0,-1,0} + 72\pi^2H_{-1,0,0,0,-1,-1} - 8\pi^2H_{-1,0,0,-1,0,0} - 24\pi^2H_{-1,0,0,-1,0,-1} \notag\\
  &\quad - 8\pi^2H_{-1,0,0,-1,-1,0} + 72\pi^2H_{-1,0,0,-1,-1,-1} - \tfrac{64}{3}\pi^2H_{-1,0,-1,0,0,0} + \tfrac{16}{3}\pi^2H_{-1,0,-1,0,-1,0} \notag\\
  &\quad - 32\pi^2H_{-1,0,-1,0,-1,-1} - \tfrac{56}{3}\pi^2H_{-1,0,-1,-1,0,0} - 8\pi^2H_{-1,0,-1,-1,0,-1} - \tfrac{16}{3}\pi^2H_{-1,0,-1,-1,-1,0} \notag\\
  &\quad + 176\pi^2H_{-1,0,-1,-1,-1,-1} - \tfrac{40}{3}\pi^2H_{-1,-1,0,0,0,0} - 8\pi^2H_{-1,-1,0,0,-1,0} + 24\pi^2H_{-1,-1,0,0,-1,-1} \notag\\
  &\quad - \tfrac{32}{3}\pi^2H_{-1,-1,0,-1,0,0} - 8\pi^2H_{-1,-1,0,-1,0,-1} - \tfrac{16}{3}\pi^2H_{-1,-1,0,-1,-1,0} + 80\pi^2H_{-1,-1,0,-1,-1,-1} \notag\\
  &\quad - \tfrac{20}{3}\pi^2H_{-1,-1,-1,0,0,0} - \tfrac{16}{3}\pi^2H_{-1,-1,-1,0,-1,0} + 32\pi^2H_{-1,-1,-1,0,-1,-1} - \tfrac{10}{3}\pi^2H_{-1,-1,-1,-1,0,0} \notag\\
  &\quad + 8\pi^2H_{-1,-1,-1,-1,0,-1} - \tfrac{2}{3}\pi^2H_{-1,-1,-1,-1,-1,0} - 2\pi^2H_{-1,-1,-1,-1,-1,-1} - 304\zeta_3H_{0,-1,0,0,0} \notag\\
  &\quad - 408\zeta_3H_{0,-1,-1,0,0} - 12\zeta_3H_{0,-1,-1,-1,0} + 372\zeta_3H_{0,-1,-1,-1,-1} - \tfrac{928}{3}\zeta_3H_{-1,0,0,0,0} \notag\\
  &\quad + 144\zeta_3H_{-1,0,0,0,-1} + 48\zeta_3H_{-1,0,0,-1,0} + 144\zeta_3H_{-1,0,0,-1,-1} - \tfrac{368}{3}\zeta_3H_{-1,0,-1,0,0} - 64\zeta_3H_{-1,0,-1,0,-1} \notag\\
  &\quad + 352\zeta_3H_{-1,0,-1,-1,-1} - \tfrac{464}{3}\zeta_3H_{-1,-1,0,0,0} + 48\zeta_3H_{-1,-1,0,0,-1} + 160\zeta_3H_{-1,-1,0,-1,-1} \notag\\
  &\quad - \tfrac{232}{3}\zeta_3H_{-1,-1,-1,0,0} + 64\zeta_3H_{-1,-1,-1,0,-1} - 36\zeta_3H_{-1,-1,-1,-1,0} - 4\zeta_3H_{-1,-1,-1,-1,-1} \notag\\
  &\quad - \tfrac{73}{30}\pi^4H_{0,-1,0,0} - \tfrac{4}{15}\pi^4H_{0,-1,-1,0} + \tfrac{131}{30}\pi^4H_{0,-1,-1,-1} - 3\pi^4H_{-1,0,0,0} + \tfrac{46}{15}\pi^4H_{-1,0,0,-1} \notag\\
  &\quad - \tfrac{2}{15}\pi^4H_{-1,0,-1,0} + \tfrac{40}{9}\pi^4H_{-1,0,-1,-1} - \tfrac{3}{2}\pi^4H_{-1,-1,0,0} + \tfrac{92}{45}\pi^4H_{-1,-1,0,-1} - \tfrac{2}{3}\pi^4H_{-1,-1,-1,0} \notag\\
  &\quad - \tfrac{17}{18}\pi^4H_{-1,-1,-1,-1} + \tfrac{58}{3}\zeta_3\pi^2H_{0,-1,0} - 208\zeta_3\pi^2H_{0,-1,-1} + \tfrac{428}{9}\zeta_3\pi^2H_{-1,0,0} - \tfrac{256}{3}\zeta_3\pi^2H_{-1,0,-1} \notag\\
  &\quad + \tfrac{82}{9}\zeta_3\pi^2H_{-1,-1,0} + \tfrac{16}{3}\zeta_3\pi^2H_{-1,-1,-1} - 36\zeta_5H_{0,-1,0} + 360\zeta_5H_{0,-1,-1} - \tfrac{2632}{5}\zeta_5H_{-1,0,0} \notag\\
  &\quad + 328\zeta_5H_{-1,0,-1} - 252\zeta_5H_{-1,-1,0} - 16\zeta_5H_{-1,-1,-1} - 390\zeta_3^2H_{0,-1} + 228\zeta_3^2H_{-1,0} + \tfrac{2}{3}\zeta_3^2H_{-1,-1} \notag\\
  &\quad - \tfrac{337}{630}\pi^6H_{0,-1} - \tfrac{577}{1260}\pi^6H_{-1,0} - \tfrac{37}{315}\pi^6H_{-1,-1} - \tfrac{59}{60}\zeta_3\pi^4H_{0} + \tfrac{2279}{270}\zeta_3\pi^4H_{-1} - \tfrac{211}{5}\zeta_5\pi^2H_{0} \notag\\
  &\quad + \tfrac{454}{5}\zeta_5\pi^2H_{-1} - \tfrac{4290}{7}\zeta_7H_{0} + 44\zeta_7H_{-1} - 144\zeta_{3,5} + \tfrac{50}{3}\zeta_3\zeta_5 + \tfrac{143}{18}\zeta_3^2\pi^2 + \tfrac{43201}{151200}\pi^8\,. \label{eq:B2}
\end{align}

\begin{align}
  c^{(3)}_{-6} &= \tfrac{16}{9}\,, \notag\\
  c^{(3)}_{-5} &= \tfrac{5}{3}H_{0}\,, \notag\\
  c^{(3)}_{-4} &= - \tfrac{3}{2}\pi^2\,, \notag\\
  c^{(3)}_{-3} &= - 3H_{-1,0,0} + \tfrac{11}{12}\pi^2H_{0} - \tfrac{3}{2}\pi^2H_{-1} - \tfrac{131}{9}\zeta_3\,, \notag\\
  c^{(3)}_{-2} &= 14H_{0,-1,0,0} - 9H_{-1,0,0,0} - 3H_{-1,-1,0,0} + 7\pi^2H_{0,-1} - \pi^2H_{-1,0} - \tfrac{3}{2}\pi^2H_{-1,-1} - \tfrac{49}{3}\zeta_3H_{0} - 3\zeta_3H_{-1} \notag\\
  &\quad - \tfrac{1411}{1080}\pi^4\,, \notag\\
  c^{(3)}_{-1} &= - 8H_{0,0,-1,0,0} + 42H_{0,-1,0,0,0} + 38H_{0,-1,-1,0,0} - 27H_{-1,0,0,0,0} + 26H_{-1,0,-1,0,0} - 9H_{-1,-1,0,0,0} \notag\\
  &\quad - 3H_{-1,-1,-1,0,0} - 4\pi^2H_{0,0,-1} + \tfrac{2}{3}\pi^2H_{0,-1,0} + 19\pi^2H_{0,-1,-1} - \tfrac{25}{4}\pi^2H_{-1,0,0} + 13\pi^2H_{-1,0,-1} \notag\\
  &\quad - \pi^2H_{-1,-1,0} - \tfrac{3}{2}\pi^2H_{-1,-1,-1} + 38\zeta_3H_{0,-1} - 58\zeta_3H_{-1,0} - 3\zeta_3H_{-1,-1} + \tfrac{503}{1440}\pi^4H_{0} - \tfrac{13}{8}\pi^4H_{-1} \notag\\
  &\quad + \tfrac{73}{4}\zeta_3\pi^2 - \tfrac{301}{15}\zeta_5\,, \notag\\
  c^{(3)}_{0} &= - 24H_{0,0,-1,0,0,0} - 56H_{0,0,-1,-1,0,0} + 126H_{0,-1,0,0,0,0} - 28H_{0,-1,0,-1,0,0} + 114H_{0,-1,-1,0,0,0} \notag\\
  &\quad + 62H_{0,-1,-1,-1,0,0} - 81H_{-1,0,0,0,0,0} + 32H_{-1,0,0,-1,0,0} + 78H_{-1,0,-1,0,0,0} + 98H_{-1,0,-1,-1,0,0} \notag\\
  &\quad - 27H_{-1,-1,0,0,0,0} + 26H_{-1,-1,0,-1,0,0} - 9H_{-1,-1,-1,0,0,0} - 3H_{-1,-1,-1,-1,0,0} + \tfrac{16}{3}\pi^2H_{0,0,-1,0} \notag\\
  &\quad - 28\pi^2H_{0,0,-1,-1} + \tfrac{13}{2}\pi^2H_{0,-1,0,0} - 14\pi^2H_{0,-1,0,-1} + \tfrac{26}{3}\pi^2H_{0,-1,-1,0} + 31\pi^2H_{0,-1,-1,-1} \notag\\
  &\quad - \tfrac{75}{4}\pi^2H_{-1,0,0,0} + 16\pi^2H_{-1,0,0,-1} - \tfrac{10}{3}\pi^2H_{-1,0,-1,0} + 49\pi^2H_{-1,0,-1,-1} - \tfrac{25}{4}\pi^2H_{-1,-1,0,0} \notag\\
  &\quad + 13\pi^2H_{-1,-1,0,-1} - \pi^2H_{-1,-1,-1,0} - \tfrac{3}{2}\pi^2H_{-1,-1,-1,-1} - 56\zeta_3H_{0,0,-1} + 108\zeta_3H_{0,-1,0} \notag\\
  &\quad + 62\zeta_3H_{0,-1,-1} - 183\zeta_3H_{-1,0,0} + 98\zeta_3H_{-1,0,-1} - 58\zeta_3H_{-1,-1,0} - 3\zeta_3H_{-1,-1,-1} + \tfrac{145}{36}\pi^4H_{0,-1} \notag\\
  &\quad - \tfrac{131}{60}\pi^4H_{-1,0} - \tfrac{13}{8}\pi^4H_{-1,-1} - \tfrac{223}{12}\zeta_3\pi^2H_{0} - \tfrac{37}{12}\zeta_3\pi^2H_{-1} - 149\zeta_5H_{0} + 57\zeta_5H_{-1} + \tfrac{167}{9}\zeta_3^2 - \tfrac{624607}{544320}\pi^6\,, \notag\\
  c^{(3)}_{1} &= - 72H_{0,0,-1,0,0,0,0} + 32H_{0,0,-1,0,-1,0,0} - 168H_{0,0,-1,-1,0,0,0} - 296H_{0,0,-1,-1,-1,0,0} + 378H_{0,-1,0,0,0,0,0} \notag\\
  &\quad + 96H_{0,-1,0,0,-1,0,0} - 84H_{0,-1,0,-1,0,0,0} - 220H_{0,-1,0,-1,-1,0,0} + 342H_{0,-1,-1,0,0,0,0} \notag\\
  &\quad - 124H_{0,-1,-1,0,-1,0,0} + 186H_{0,-1,-1,-1,0,0,0} + 86H_{0,-1,-1,-1,-1,0,0} - 243H_{-1,0,0,0,0,0,0} \notag\\
  &\quad + 136H_{-1,0,0,0,-1,0,0} + 96H_{-1,0,0,-1,0,0,0} + 176H_{-1,0,0,-1,-1,0,0} + 234H_{-1,0,-1,0,0,0,0} \notag\\
  &\quad - 36H_{-1,0,-1,0,-1,0,0} + 294H_{-1,0,-1,-1,0,0,0} + 170H_{-1,0,-1,-1,-1,0,0} - 81H_{-1,-1,0,0,0,0,0} \notag\\
  &\quad + 32H_{-1,-1,0,0,-1,0,0} + 78H_{-1,-1,0,-1,0,0,0} + 98H_{-1,-1,0,-1,-1,0,0} - 27H_{-1,-1,-1,0,0,0,0} \notag\\
  &\quad + 26H_{-1,-1,-1,0,-1,0,0} - 9H_{-1,-1,-1,-1,0,0,0} - 3H_{-1,-1,-1,-1,-1,0,0} + 18\pi^2H_{0,0,-1,0,0} \notag\\
  &\quad + 16\pi^2H_{0,0,-1,0,-1} + \tfrac{64}{3}\pi^2H_{0,0,-1,-1,0} - 148\pi^2H_{0,0,-1,-1,-1} + \tfrac{39}{2}\pi^2H_{0,-1,0,0,0} + 48\pi^2H_{0,-1,0,0,-1} \notag\\
  &\quad + \tfrac{68}{3}\pi^2H_{0,-1,0,-1,0} - 110\pi^2H_{0,-1,0,-1,-1} + \tfrac{81}{2}\pi^2H_{0,-1,-1,0,0} - 62\pi^2H_{0,-1,-1,0,-1} \notag\\
  &\quad + \tfrac{50}{3}\pi^2H_{0,-1,-1,-1,0} + 43\pi^2H_{0,-1,-1,-1,-1} - \tfrac{225}{4}\pi^2H_{-1,0,0,0,0} + 68\pi^2H_{-1,0,0,0,-1} - \tfrac{40}{3}\pi^2H_{-1,0,0,-1,0} \notag\\
  &\quad + 88\pi^2H_{-1,0,0,-1,-1} - \tfrac{1}{2}\pi^2H_{-1,0,-1,0,0} - 18\pi^2H_{-1,0,-1,0,-1} + \tfrac{62}{3}\pi^2H_{-1,0,-1,-1,0} \notag\\
  &\quad + 85\pi^2H_{-1,0,-1,-1,-1} - \tfrac{75}{4}\pi^2H_{-1,-1,0,0,0} + 16\pi^2H_{-1,-1,0,0,-1} - \tfrac{10}{3}\pi^2H_{-1,-1,0,-1,0} \notag\\
  &\quad + 49\pi^2H_{-1,-1,0,-1,-1} - \tfrac{25}{4}\pi^2H_{-1,-1,-1,0,0} + 13\pi^2H_{-1,-1,-1,0,-1} - \pi^2H_{-1,-1,-1,-1,0} \notag\\
  &\quad - \tfrac{3}{2}\pi^2H_{-1,-1,-1,-1,-1} + 64\zeta_3H_{0,0,-1,0} - 296\zeta_3H_{0,0,-1,-1} + 110\zeta_3H_{0,-1,0,0} - 220\zeta_3H_{0,-1,0,-1} \notag\\
  &\quad + 348\zeta_3H_{0,-1,-1,0} + 86\zeta_3H_{0,-1,-1,-1} - 549\zeta_3H_{-1,0,0,0} + 176\zeta_3H_{-1,0,0,-1} + 196\zeta_3H_{-1,0,-1,0} \notag\\
  &\quad + 170\zeta_3H_{-1,0,-1,-1} - 183\zeta_3H_{-1,-1,0,0} + 98\zeta_3H_{-1,-1,0,-1} - 58\zeta_3H_{-1,-1,-1,0} - 3\zeta_3H_{-1,-1,-1,-1} \notag\\
  &\quad - \tfrac{31}{9}\pi^4H_{0,0,-1} - \tfrac{173}{90}\pi^4H_{0,-1,0} + \tfrac{2297}{180}\pi^4H_{0,-1,-1} - \tfrac{1443}{160}\pi^4H_{-1,0,0} + \tfrac{1271}{180}\pi^4H_{-1,0,-1} - \tfrac{131}{60}\pi^4H_{-1,-1,0} \notag\\
  &\quad - \tfrac{13}{8}\pi^4H_{-1,-1,-1} - \tfrac{257}{2}\zeta_3\pi^2H_{0,-1} - \tfrac{101}{2}\zeta_3\pi^2H_{-1,0} - \tfrac{37}{12}\zeta_3\pi^2H_{-1,-1} + 262\zeta_5H_{0,-1} - 566\zeta_5H_{-1,0} \notag\\
  &\quad + 57\zeta_5H_{-1,-1} - \tfrac{211}{6}\zeta_3^2H_{0} - 155\zeta_3^2H_{-1} - \tfrac{21853}{51840}\pi^6H_{0} - \tfrac{85567}{60480}\pi^6H_{-1} + \tfrac{76619}{4320}\zeta_3\pi^4 + \tfrac{4313}{20}\zeta_5\pi^2 - \tfrac{1865}{7}\zeta_7\,, \notag\\
  c^{(3)}_{2} &= - 216H_{0,0,-1,0,0,0,0,0} + 96H_{0,0,-1,0,0,-1,0,0} + 96H_{0,0,-1,0,-1,0,0,0} + 32H_{0,0,-1,0,-1,-1,0,0} \notag\\
  &\quad - 504H_{0,0,-1,-1,0,0,0,0} + 32H_{0,0,-1,-1,0,-1,0,0} - 888H_{0,0,-1,-1,-1,0,0,0} - 1208H_{0,0,-1,-1,-1,-1,0,0} \notag\\
  &\quad + 1134H_{0,-1,0,0,0,0,0,0} + 288H_{0,-1,0,0,0,-1,0,0} + 288H_{0,-1,0,0,-1,0,0,0} + 192H_{0,-1,0,0,-1,-1,0,0} \notag\\
  &\quad - 252H_{0,-1,0,-1,0,0,0,0} + 184H_{0,-1,0,-1,0,-1,0,0} - 660H_{0,-1,0,-1,-1,0,0,0} - 1084H_{0,-1,0,-1,-1,-1,0,0} \notag\\
  &\quad + 1026H_{0,-1,-1,0,0,0,0,0} + 96H_{0,-1,-1,0,0,-1,0,0} - 372H_{0,-1,-1,0,-1,0,0,0} - 412H_{0,-1,-1,0,-1,-1,0,0} \notag\\
  &\quad + 558H_{0,-1,-1,-1,0,0,0,0} - 220H_{0,-1,-1,-1,0,-1,0,0} + 258H_{0,-1,-1,-1,-1,0,0,0} + 110H_{0,-1,-1,-1,-1,-1,0,0} \notag\\
  &\quad - 729H_{-1,0,0,0,0,0,0,0} + 408H_{-1,0,0,0,0,-1,0,0} + 408H_{-1,0,0,0,-1,0,0,0} + 1000H_{-1,0,0,0,-1,-1,0,0} \notag\\
  &\quad + 288H_{-1,0,0,-1,0,0,0,0} + 176H_{-1,0,0,-1,0,-1,0,0} + 528H_{-1,0,0,-1,-1,0,0,0} + 32H_{-1,0,0,-1,-1,-1,0,0} \notag\\
  &\quad + 702H_{-1,0,-1,0,0,0,0,0} + 240H_{-1,0,-1,0,0,-1,0,0} - 108H_{-1,0,-1,0,-1,0,0,0} - 516H_{-1,0,-1,0,-1,-1,0,0} \notag\\
  &\quad + 882H_{-1,0,-1,-1,0,0,0,0} - 324H_{-1,0,-1,-1,0,-1,0,0} + 510H_{-1,0,-1,-1,-1,0,0,0} + 242H_{-1,0,-1,-1,-1,-1,0,0} \notag\\
  &\quad - 243H_{-1,-1,0,0,0,0,0,0} + 136H_{-1,-1,0,0,0,-1,0,0} + 96H_{-1,-1,0,0,-1,0,0,0} + 176H_{-1,-1,0,0,-1,-1,0,0} \notag\\
  &\quad + 234H_{-1,-1,0,-1,0,0,0,0} - 36H_{-1,-1,0,-1,0,-1,0,0} + 294H_{-1,-1,0,-1,-1,0,0,0} + 170H_{-1,-1,0,-1,-1,-1,0,0} \notag\\
  &\quad - 81H_{-1,-1,-1,0,0,0,0,0} + 32H_{-1,-1,-1,0,0,-1,0,0} + 78H_{-1,-1,-1,0,-1,0,0,0} + 98H_{-1,-1,-1,0,-1,-1,0,0} \notag\\
  &\quad - 27H_{-1,-1,-1,-1,0,0,0,0} + 26H_{-1,-1,-1,-1,0,-1,0,0} - 9H_{-1,-1,-1,-1,-1,0,0,0} - 3H_{-1,-1,-1,-1,-1,-1,0,0} \notag\\
  &\quad + 54\pi^2H_{0,0,-1,0,0,0} + 48\pi^2H_{0,0,-1,0,0,-1} + \tfrac{32}{3}\pi^2H_{0,0,-1,0,-1,0} + 16\pi^2H_{0,0,-1,0,-1,-1} \notag\\
  &\quad + 78\pi^2H_{0,0,-1,-1,0,0} + 16\pi^2H_{0,0,-1,-1,0,-1} + \tfrac{160}{3}\pi^2H_{0,0,-1,-1,-1,0} - 604\pi^2H_{0,0,-1,-1,-1,-1} \notag\\
  &\quad + \tfrac{117}{2}\pi^2H_{0,-1,0,0,0,0} + 144\pi^2H_{0,-1,0,0,0,-1} + 16\pi^2H_{0,-1,0,0,-1,0} + 96\pi^2H_{0,-1,0,0,-1,-1} \notag\\
  &\quad + 59\pi^2H_{0,-1,0,-1,0,0} + 92\pi^2H_{0,-1,0,-1,0,-1} + \tfrac{212}{3}\pi^2H_{0,-1,0,-1,-1,0} - 542\pi^2H_{0,-1,0,-1,-1,-1} \notag\\
  &\quad + \tfrac{243}{2}\pi^2H_{0,-1,-1,0,0,0} + 48\pi^2H_{0,-1,-1,0,0,-1} + \tfrac{20}{3}\pi^2H_{0,-1,-1,0,-1,0} - 206\pi^2H_{0,-1,-1,0,-1,-1} \notag\\
  &\quad + \tfrac{149}{2}\pi^2H_{0,-1,-1,-1,0,0} - 110\pi^2H_{0,-1,-1,-1,0,-1} + \tfrac{74}{3}\pi^2H_{0,-1,-1,-1,-1,0} + 55\pi^2H_{0,-1,-1,-1,-1,-1} \notag\\
  &\quad - \tfrac{675}{4}\pi^2H_{-1,0,0,0,0,0} + 204\pi^2H_{-1,0,0,0,0,-1} - \tfrac{296}{3}\pi^2H_{-1,0,0,0,-1,0} + 500\pi^2H_{-1,0,0,0,-1,-1} \notag\\
  &\quad - 40\pi^2H_{-1,0,0,-1,0,0} + 88\pi^2H_{-1,0,0,-1,0,-1} + \tfrac{248}{3}\pi^2H_{-1,0,0,-1,-1,0} + 16\pi^2H_{-1,0,0,-1,-1,-1} \notag\\
  &\quad - \tfrac{3}{2}\pi^2H_{-1,0,-1,0,0,0} + 120\pi^2H_{-1,0,-1,0,0,-1} + 68\pi^2H_{-1,0,-1,0,-1,0} - 258\pi^2H_{-1,0,-1,0,-1,-1} \notag\\
  &\quad + \tfrac{203}{2}\pi^2H_{-1,0,-1,-1,0,0} - 162\pi^2H_{-1,0,-1,-1,0,-1} + \tfrac{134}{3}\pi^2H_{-1,0,-1,-1,-1,0} + 121\pi^2H_{-1,0,-1,-1,-1,-1} \notag\\
  &\quad - \tfrac{225}{4}\pi^2H_{-1,-1,0,0,0,0} + 68\pi^2H_{-1,-1,0,0,0,-1} - \tfrac{40}{3}\pi^2H_{-1,-1,0,0,-1,0} + 88\pi^2H_{-1,-1,0,0,-1,-1} \notag\\
  &\quad - \tfrac{1}{2}\pi^2H_{-1,-1,0,-1,0,0} - 18\pi^2H_{-1,-1,0,-1,0,-1} + \tfrac{62}{3}\pi^2H_{-1,-1,0,-1,-1,0} + 85\pi^2H_{-1,-1,0,-1,-1,-1} \notag\\
  &\quad - \tfrac{75}{4}\pi^2H_{-1,-1,-1,0,0,0} + 16\pi^2H_{-1,-1,-1,0,0,-1} - \tfrac{10}{3}\pi^2H_{-1,-1,-1,0,-1,0} + 49\pi^2H_{-1,-1,-1,0,-1,-1} \notag\\
  &\quad - \tfrac{25}{4}\pi^2H_{-1,-1,-1,-1,0,0} + 13\pi^2H_{-1,-1,-1,-1,0,-1} - \pi^2H_{-1,-1,-1,-1,-1,0} - \tfrac{3}{2}\pi^2H_{-1,-1,-1,-1,-1,-1} \notag\\
  &\quad + 424\zeta_3H_{0,0,-1,0,0} + 32\zeta_3H_{0,0,-1,0,-1} + 256\zeta_3H_{0,0,-1,-1,0} - 1208\zeta_3H_{0,0,-1,-1,-1} + 330\zeta_3H_{0,-1,0,0,0} \notag\\
  &\quad + 192\zeta_3H_{0,-1,0,0,-1} + 56\zeta_3H_{0,-1,0,-1,0} - 1084\zeta_3H_{0,-1,0,-1,-1} + 518\zeta_3H_{0,-1,-1,0,0} \notag\\
  &\quad - 412\zeta_3H_{0,-1,-1,0,-1} + 588\zeta_3H_{0,-1,-1,-1,0} + 110\zeta_3H_{0,-1,-1,-1,-1} - 1647\zeta_3H_{-1,0,0,0,0} \notag\\
  &\quad + 1000\zeta_3H_{-1,0,0,0,-1} + 144\zeta_3H_{-1,0,0,-1,0} + 32\zeta_3H_{-1,0,0,-1,-1} + 218\zeta_3H_{-1,0,-1,0,0} \notag\\
  &\quad - 516\zeta_3H_{-1,0,-1,0,-1} + 916\zeta_3H_{-1,0,-1,-1,0} + 242\zeta_3H_{-1,0,-1,-1,-1} - 549\zeta_3H_{-1,-1,0,0,0} \notag\\
  &\quad + 176\zeta_3H_{-1,-1,0,0,-1} + 196\zeta_3H_{-1,-1,0,-1,0} + 170\zeta_3H_{-1,-1,0,-1,-1} - 183\zeta_3H_{-1,-1,-1,0,0} \notag\\
  &\quad + 98\zeta_3H_{-1,-1,-1,0,-1} - 58\zeta_3H_{-1,-1,-1,-1,0} - 3\zeta_3H_{-1,-1,-1,-1,-1} - \tfrac{32}{45}\pi^4H_{0,0,-1,0} - \tfrac{689}{45}\pi^4H_{0,0,-1,-1} \notag\\
  &\quad - \tfrac{1691}{240}\pi^4H_{0,-1,0,0} - \tfrac{1469}{90}\pi^4H_{0,-1,0,-1} + \tfrac{223}{90}\pi^4H_{0,-1,-1,0} + \tfrac{3869}{180}\pi^4H_{0,-1,-1,-1} - \tfrac{4329}{160}\pi^4H_{-1,0,0,0} \notag\\
  &\quad - \tfrac{76}{45}\pi^4H_{-1,0,0,-1} - \tfrac{139}{18}\pi^4H_{-1,0,-1,0} + \tfrac{5987}{180}\pi^4H_{-1,0,-1,-1} - \tfrac{1443}{160}\pi^4H_{-1,-1,0,0} + \tfrac{1271}{180}\pi^4H_{-1,-1,0,-1} \notag\\
  &\quad - \tfrac{131}{60}\pi^4H_{-1,-1,-1,0} - \tfrac{13}{8}\pi^4H_{-1,-1,-1,-1} + \tfrac{358}{3}\zeta_3\pi^2H_{0,0,-1} - \tfrac{463}{3}\zeta_3\pi^2H_{0,-1,0} - \tfrac{453}{2}\zeta_3\pi^2H_{0,-1,-1} \notag\\
  &\quad - \tfrac{453}{4}\zeta_3\pi^2H_{-1,0,0} - \tfrac{1585}{6}\zeta_3\pi^2H_{-1,0,-1} - \tfrac{101}{2}\zeta_3\pi^2H_{-1,-1,0} - \tfrac{37}{12}\zeta_3\pi^2H_{-1,-1,-1} - 760\zeta_5H_{0,0,-1} \notag\\
  &\quad + 1428\zeta_5H_{0,-1,0} + 910\zeta_5H_{0,-1,-1} - \tfrac{8301}{5}\zeta_5H_{-1,0,0} + 178\zeta_5H_{-1,0,-1} - 566\zeta_5H_{-1,-1,0} \notag\\
  &\quad + 57\zeta_5H_{-1,-1,-1} - 162\zeta_3^2H_{0,-1} - 204\zeta_3^2H_{-1,0} - 155\zeta_3^2H_{-1,-1} + \tfrac{30001}{10080}\pi^6H_{0,-1} - \tfrac{85769}{30240}\pi^6H_{-1,0} \notag\\
  &\quad - \tfrac{85567}{60480}\pi^6H_{-1,-1} - \tfrac{20827}{1440}\zeta_3\pi^4H_{0} + \tfrac{8449}{1440}\zeta_3\pi^4H_{-1} - \tfrac{4391}{20}\zeta_5\pi^2H_{0} + \tfrac{8399}{60}\zeta_5\pi^2H_{-1} - \tfrac{13585}{7}\zeta_7H_{0} \notag\\
  &\quad + 1275\zeta_7H_{-1} - \tfrac{512}{5}\zeta_{3,5} - \tfrac{26824}{15}\zeta_3\zeta_5 - \tfrac{241}{2}\zeta_3^2\pi^2 - \tfrac{7760099}{18144000}\pi^8\,. \label{eq:B3}
\end{align}

\begin{align}
  c^{(4)}_{-8} &= \tfrac{4}{9}\,, \notag\\
  c^{(4)}_{-7} &= \tfrac{35}{72}H_{0}\,, \notag\\
  c^{(4)}_{-6} &= - \tfrac{187}{432}\pi^2\,, \notag\\
  c^{(4)}_{-5} &= - \tfrac{10}{9}H_{-1,0,0} + \tfrac{115}{216}\pi^2H_{0} - \tfrac{5}{9}\pi^2H_{-1} - \tfrac{1169}{216}\zeta_3\,, \notag\\
  c^{(4)}_{-4} &= \tfrac{20}{3}H_{0,-1,0,0} - \tfrac{40}{9}H_{-1,0,0,0} - \tfrac{10}{9}H_{-1,-1,0,0} + \tfrac{10}{3}\pi^2H_{0,-1} - \tfrac{5}{9}\pi^2H_{-1,0} - \tfrac{5}{9}\pi^2H_{-1,-1} - \tfrac{485}{54}\zeta_3H_{0} \notag\\
  &\quad - \tfrac{10}{9}\zeta_3H_{-1} - \tfrac{2339}{3240}\pi^4\,, \notag\\
  c^{(4)}_{-3} &= - \tfrac{20}{3}H_{0,0,-1,0,0} + \tfrac{80}{3}H_{0,-1,0,0,0} + \tfrac{50}{3}H_{0,-1,-1,0,0} - \tfrac{160}{9}H_{-1,0,0,0,0} + \tfrac{44}{3}H_{-1,0,-1,0,0} - \tfrac{40}{9}H_{-1,-1,0,0,0} \notag\\
  &\quad - \tfrac{10}{9}H_{-1,-1,-1,0,0} - \tfrac{10}{3}\pi^2H_{0,0,-1} + \tfrac{5}{3}\pi^2H_{0,-1,0} + \tfrac{25}{3}\pi^2H_{0,-1,-1} - \tfrac{116}{27}\pi^2H_{-1,0,0} + \tfrac{22}{3}\pi^2H_{-1,0,-1} \notag\\
  &\quad - \tfrac{5}{9}\pi^2H_{-1,-1,0} - \tfrac{5}{9}\pi^2H_{-1,-1,-1} + \tfrac{50}{3}\zeta_3H_{0,-1} - \tfrac{98}{3}\zeta_3H_{-1,0} - \tfrac{10}{9}\zeta_3H_{-1,-1} + \tfrac{371}{648}\pi^4H_{0} - \tfrac{23}{20}\pi^4H_{-1} \notag\\
  &\quad + \tfrac{7985}{648}\zeta_3\pi^2 + \tfrac{673}{40}\zeta_5\,, \notag\\
  c^{(4)}_{-2} &= \tfrac{14}{3}H_{0,0,0,-1,0,0} - \tfrac{80}{3}H_{0,0,-1,0,0,0} - \tfrac{140}{3}H_{0,0,-1,-1,0,0} + \tfrac{320}{3}H_{0,-1,0,0,0,0} - 36H_{0,-1,0,-1,0,0} \notag\\
  &\quad + \tfrac{200}{3}H_{0,-1,-1,0,0,0} + \tfrac{50}{3}H_{0,-1,-1,-1,0,0} - \tfrac{640}{9}H_{-1,0,0,0,0,0} + \tfrac{88}{3}H_{-1,0,0,-1,0,0} + \tfrac{176}{3}H_{-1,0,-1,0,0,0} \notag\\
  &\quad + \tfrac{164}{3}H_{-1,0,-1,-1,0,0} - \tfrac{160}{9}H_{-1,-1,0,0,0,0} + \tfrac{44}{3}H_{-1,-1,0,-1,0,0} - \tfrac{40}{9}H_{-1,-1,-1,0,0,0} - \tfrac{10}{9}H_{-1,-1,-1,-1,0,0} \notag\\
  &\quad + \tfrac{7}{3}\pi^2H_{0,0,0,-1} + \tfrac{10}{3}\pi^2H_{0,0,-1,0} - \tfrac{70}{3}\pi^2H_{0,0,-1,-1} + \tfrac{118}{9}\pi^2H_{0,-1,0,0} - 18\pi^2H_{0,-1,0,-1} \notag\\
  &\quad + \tfrac{25}{3}\pi^2H_{0,-1,-1,0} + \tfrac{25}{3}\pi^2H_{0,-1,-1,-1} - \tfrac{464}{27}\pi^2H_{-1,0,0,0} + \tfrac{44}{3}\pi^2H_{-1,0,0,-1} + \tfrac{2}{3}\pi^2H_{-1,0,-1,0} \notag\\
  &\quad + \tfrac{82}{3}\pi^2H_{-1,0,-1,-1} - \tfrac{116}{27}\pi^2H_{-1,-1,0,0} + \tfrac{22}{3}\pi^2H_{-1,-1,0,-1} - \tfrac{5}{9}\pi^2H_{-1,-1,-1,0} - \tfrac{5}{9}\pi^2H_{-1,-1,-1,-1} \notag\\
  &\quad - \tfrac{140}{3}\zeta_3H_{0,0,-1} + 122\zeta_3H_{0,-1,0} + \tfrac{50}{3}\zeta_3H_{0,-1,-1} - \tfrac{3668}{27}\zeta_3H_{-1,0,0} + \tfrac{164}{3}\zeta_3H_{-1,0,-1} - \tfrac{98}{3}\zeta_3H_{-1,-1,0} \notag\\
  &\quad - \tfrac{10}{9}\zeta_3H_{-1,-1,-1} + \tfrac{263}{60}\pi^4H_{0,-1} - \tfrac{259}{135}\pi^4H_{-1,0} - \tfrac{23}{20}\pi^4H_{-1,-1} - \tfrac{2635}{162}\zeta_3\pi^2H_{0} - \tfrac{23}{3}\zeta_3\pi^2H_{-1} \notag\\
  &\quad - \tfrac{2377}{18}\zeta_5H_{0} + \tfrac{344}{9}\zeta_5H_{-1} + \tfrac{997}{162}\zeta_3^2 - \tfrac{10867}{7560}\pi^6\,, \notag\\
  c^{(4)}_{-1} &= \tfrac{16}{3}H_{0,0,0,0,-1,0,0} + \tfrac{56}{3}H_{0,0,0,-1,0,0,0} + \tfrac{140}{3}H_{0,0,0,-1,-1,0,0} - \tfrac{320}{3}H_{0,0,-1,0,0,0,0} + 32H_{0,0,-1,0,-1,0,0} \notag\\
  &\quad - \tfrac{560}{3}H_{0,0,-1,-1,0,0,0} - \tfrac{668}{3}H_{0,0,-1,-1,-1,0,0} + \tfrac{1280}{3}H_{0,-1,0,0,0,0,0} + 48H_{0,-1,0,0,-1,0,0} \notag\\
  &\quad - 144H_{0,-1,0,-1,0,0,0} - 252H_{0,-1,0,-1,-1,0,0} + \tfrac{800}{3}H_{0,-1,-1,0,0,0,0} - 148H_{0,-1,-1,0,-1,0,0} \notag\\
  &\quad + \tfrac{200}{3}H_{0,-1,-1,-1,0,0,0} + \tfrac{50}{3}H_{0,-1,-1,-1,-1,0,0} - \tfrac{2560}{9}H_{-1,0,0,0,0,0,0} + \tfrac{488}{3}H_{-1,0,0,0,-1,0,0} \notag\\
  &\quad + \tfrac{352}{3}H_{-1,0,0,-1,0,0,0} + \tfrac{400}{3}H_{-1,0,0,-1,-1,0,0} + \tfrac{704}{3}H_{-1,0,-1,0,0,0,0} - 32H_{-1,0,-1,0,-1,0,0} \notag\\
  &\quad + \tfrac{656}{3}H_{-1,0,-1,-1,0,0,0} + \tfrac{164}{3}H_{-1,0,-1,-1,-1,0,0} - \tfrac{640}{9}H_{-1,-1,0,0,0,0,0} + \tfrac{88}{3}H_{-1,-1,0,0,-1,0,0} \notag\\
  &\quad + \tfrac{176}{3}H_{-1,-1,0,-1,0,0,0} + \tfrac{164}{3}H_{-1,-1,0,-1,-1,0,0} - \tfrac{160}{9}H_{-1,-1,-1,0,0,0,0} + \tfrac{44}{3}H_{-1,-1,-1,0,-1,0,0} \notag\\
  &\quad - \tfrac{40}{9}H_{-1,-1,-1,-1,0,0,0} - \tfrac{10}{9}H_{-1,-1,-1,-1,-1,0,0} + \tfrac{8}{3}\pi^2H_{0,0,0,0,-1} - \tfrac{14}{3}\pi^2H_{0,0,0,-1,0} + \tfrac{70}{3}\pi^2H_{0,0,0,-1,-1} \notag\\
  &\quad + \tfrac{116}{9}\pi^2H_{0,0,-1,0,0} + 16\pi^2H_{0,0,-1,0,-1} + 6\pi^2H_{0,0,-1,-1,0} - \tfrac{334}{3}\pi^2H_{0,0,-1,-1,-1} + \tfrac{472}{9}\pi^2H_{0,-1,0,0,0} \notag\\
  &\quad + 24\pi^2H_{0,-1,0,0,-1} + 18\pi^2H_{0,-1,0,-1,0} - 126\pi^2H_{0,-1,0,-1,-1} + \tfrac{472}{9}\pi^2H_{0,-1,-1,0,0} - 74\pi^2H_{0,-1,-1,0,-1} \notag\\
  &\quad + \tfrac{25}{3}\pi^2H_{0,-1,-1,-1,0} + \tfrac{25}{3}\pi^2H_{0,-1,-1,-1,-1} - \tfrac{1856}{27}\pi^2H_{-1,0,0,0,0} + \tfrac{244}{3}\pi^2H_{-1,0,0,0,-1} - \tfrac{8}{3}\pi^2H_{-1,0,0,-1,0} \notag\\
  &\quad + \tfrac{200}{3}\pi^2H_{-1,0,0,-1,-1} + \tfrac{148}{9}\pi^2H_{-1,0,-1,0,0} - 16\pi^2H_{-1,0,-1,0,-1} + \tfrac{82}{3}\pi^2H_{-1,0,-1,-1,0} \notag\\
  &\quad + \tfrac{82}{3}\pi^2H_{-1,0,-1,-1,-1} - \tfrac{464}{27}\pi^2H_{-1,-1,0,0,0} + \tfrac{44}{3}\pi^2H_{-1,-1,0,0,-1} + \tfrac{2}{3}\pi^2H_{-1,-1,0,-1,0} \notag\\
  &\quad + \tfrac{82}{3}\pi^2H_{-1,-1,0,-1,-1} - \tfrac{116}{27}\pi^2H_{-1,-1,-1,0,0} + \tfrac{22}{3}\pi^2H_{-1,-1,-1,0,-1} - \tfrac{5}{9}\pi^2H_{-1,-1,-1,-1,0} \notag\\
  &\quad - \tfrac{5}{9}\pi^2H_{-1,-1,-1,-1,-1} + \tfrac{140}{3}\zeta_3H_{0,0,0,-1} - 28\zeta_3H_{0,0,-1,0} - \tfrac{668}{3}\zeta_3H_{0,0,-1,-1} + \tfrac{3052}{9}\zeta_3H_{0,-1,0,0} \notag\\
  &\quad - 252\zeta_3H_{0,-1,0,-1} + 346\zeta_3H_{0,-1,-1,0} + \tfrac{50}{3}\zeta_3H_{0,-1,-1,-1} - \tfrac{14672}{27}\zeta_3H_{-1,0,0,0} + \tfrac{400}{3}\zeta_3H_{-1,0,0,-1} \notag\\
  &\quad + 228\zeta_3H_{-1,0,-1,0} + \tfrac{164}{3}\zeta_3H_{-1,0,-1,-1} - \tfrac{3668}{27}\zeta_3H_{-1,-1,0,0} + \tfrac{164}{3}\zeta_3H_{-1,-1,0,-1} - \tfrac{98}{3}\zeta_3H_{-1,-1,-1,0} \notag\\
  &\quad - \tfrac{10}{9}\zeta_3H_{-1,-1,-1,-1} - \tfrac{283}{90}\pi^4H_{0,0,-1} + \tfrac{13}{15}\pi^4H_{0,-1,0} + \tfrac{273}{20}\pi^4H_{0,-1,-1} - \tfrac{4394}{405}\pi^4H_{-1,0,0} \notag\\
  &\quad + \tfrac{263}{30}\pi^4H_{-1,0,-1} - \tfrac{259}{135}\pi^4H_{-1,-1,0} - \tfrac{23}{20}\pi^4H_{-1,-1,-1} - \tfrac{199}{3}\zeta_3\pi^2H_{0,-1} - \tfrac{2134}{27}\zeta_3\pi^2H_{-1,0} \notag\\
  &\quad - \tfrac{23}{3}\zeta_3\pi^2H_{-1,-1} + \tfrac{272}{3}\zeta_5H_{0,-1} - \tfrac{1490}{3}\zeta_5H_{-1,0} + \tfrac{344}{9}\zeta_5H_{-1,-1} - \tfrac{1090}{81}\zeta_3^2H_{0} - \tfrac{5135}{27}\zeta_3^2H_{-1} + \tfrac{127}{1134}\pi^6H_{0} \notag\\
  &\quad - \tfrac{859}{405}\pi^6H_{-1} + \tfrac{210521}{9720}\zeta_3\pi^4 + \tfrac{28151}{120}\zeta_5\pi^2 + \tfrac{6809}{504}\zeta_7\,, \notag\\
  c^{(4)}_{0} &= - \tfrac{16}{3}H_{0,0,0,0,0,-1,0,0} + \tfrac{64}{3}H_{0,0,0,0,-1,0,0,0} + \tfrac{160}{3}H_{0,0,0,0,-1,-1,0,0} + \tfrac{224}{3}H_{0,0,0,-1,0,0,0,0} \notag\\
  &\quad - 34H_{0,0,0,-1,0,-1,0,0} + \tfrac{560}{3}H_{0,0,0,-1,-1,0,0,0} + \tfrac{1112}{3}H_{0,0,0,-1,-1,-1,0,0} - \tfrac{1280}{3}H_{0,0,-1,0,0,0,0,0} \notag\\
  &\quad - 32H_{0,0,-1,0,0,-1,0,0} + 128H_{0,0,-1,0,-1,0,0,0} + 144H_{0,0,-1,0,-1,-1,0,0} - \tfrac{2240}{3}H_{0,0,-1,-1,0,0,0,0} \notag\\
  &\quad + 48H_{0,0,-1,-1,0,-1,0,0} - \tfrac{2672}{3}H_{0,0,-1,-1,-1,0,0,0} - \tfrac{2060}{3}H_{0,0,-1,-1,-1,-1,0,0} + \tfrac{5120}{3}H_{0,-1,0,0,0,0,0,0} \notag\\
  &\quad + 112H_{0,-1,0,0,0,-1,0,0} + 192H_{0,-1,0,0,-1,0,0,0} - 72H_{0,-1,0,0,-1,-1,0,0} - 576H_{0,-1,0,-1,0,0,0,0} \notag\\
  &\quad + 232H_{0,-1,0,-1,0,-1,0,0} - 1008H_{0,-1,0,-1,-1,0,0,0} - 1020H_{0,-1,0,-1,-1,-1,0,0} + \tfrac{3200}{3}H_{0,-1,-1,0,0,0,0,0} \notag\\
  &\quad - 128H_{0,-1,-1,0,0,-1,0,0} - 592H_{0,-1,-1,0,-1,0,0,0} - 388H_{0,-1,-1,0,-1,-1,0,0} + \tfrac{800}{3}H_{0,-1,-1,-1,0,0,0,0} \notag\\
  &\quad - 148H_{0,-1,-1,-1,0,-1,0,0} + \tfrac{200}{3}H_{0,-1,-1,-1,-1,0,0,0} + \tfrac{50}{3}H_{0,-1,-1,-1,-1,-1,0,0} - \tfrac{10240}{9}H_{-1,0,0,0,0,0,0,0} \notag\\
  &\quad + \tfrac{1888}{3}H_{-1,0,0,0,0,-1,0,0} + \tfrac{1952}{3}H_{-1,0,0,0,-1,0,0,0} + \tfrac{3224}{3}H_{-1,0,0,0,-1,-1,0,0} + \tfrac{1408}{3}H_{-1,0,0,-1,0,0,0,0} \notag\\
  &\quad + 328H_{-1,0,0,-1,0,-1,0,0} + \tfrac{1600}{3}H_{-1,0,0,-1,-1,0,0,0} - \tfrac{848}{3}H_{-1,0,0,-1,-1,-1,0,0} + \tfrac{2816}{3}H_{-1,0,-1,0,0,0,0,0} \notag\\
  &\quad + 336H_{-1,0,-1,0,0,-1,0,0} - 128H_{-1,0,-1,0,-1,0,0,0} - 632H_{-1,0,-1,0,-1,-1,0,0} + \tfrac{2624}{3}H_{-1,0,-1,-1,0,0,0,0} \notag\\
  &\quad - 480H_{-1,0,-1,-1,0,-1,0,0} + \tfrac{656}{3}H_{-1,0,-1,-1,-1,0,0,0} + \tfrac{164}{3}H_{-1,0,-1,-1,-1,-1,0,0} - \tfrac{2560}{9}H_{-1,-1,0,0,0,0,0,0} \notag\\
  &\quad + \tfrac{488}{3}H_{-1,-1,0,0,0,-1,0,0} + \tfrac{352}{3}H_{-1,-1,0,0,-1,0,0,0} + \tfrac{400}{3}H_{-1,-1,0,0,-1,-1,0,0} + \tfrac{704}{3}H_{-1,-1,0,-1,0,0,0,0} \notag\\
  &\quad - 32H_{-1,-1,0,-1,0,-1,0,0} + \tfrac{656}{3}H_{-1,-1,0,-1,-1,0,0,0} + \tfrac{164}{3}H_{-1,-1,0,-1,-1,-1,0,0} - \tfrac{640}{9}H_{-1,-1,-1,0,0,0,0,0} \notag\\
  &\quad + \tfrac{88}{3}H_{-1,-1,-1,0,0,-1,0,0} + \tfrac{176}{3}H_{-1,-1,-1,0,-1,0,0,0} + \tfrac{164}{3}H_{-1,-1,-1,0,-1,-1,0,0} - \tfrac{160}{9}H_{-1,-1,-1,-1,0,0,0,0} \notag\\
  &\quad + \tfrac{44}{3}H_{-1,-1,-1,-1,0,-1,0,0} - \tfrac{40}{9}H_{-1,-1,-1,-1,-1,0,0,0} - \tfrac{10}{9}H_{-1,-1,-1,-1,-1,-1,0,0} - \tfrac{8}{3}\pi^2H_{0,0,0,0,0,-1} \notag\\
  &\quad - \tfrac{16}{3}\pi^2H_{0,0,0,0,-1,0} + \tfrac{80}{3}\pi^2H_{0,0,0,0,-1,-1} - \tfrac{158}{9}\pi^2H_{0,0,0,-1,0,0} - 17\pi^2H_{0,0,0,-1,0,-1} \notag\\
  &\quad - \tfrac{92}{3}\pi^2H_{0,0,0,-1,-1,0} + \tfrac{556}{3}\pi^2H_{0,0,0,-1,-1,-1} + \tfrac{464}{9}\pi^2H_{0,0,-1,0,0,0} - 16\pi^2H_{0,0,-1,0,0,-1} \notag\\
  &\quad - \tfrac{8}{3}\pi^2H_{0,0,-1,0,-1,0} + 72\pi^2H_{0,0,-1,0,-1,-1} + \tfrac{188}{9}\pi^2H_{0,0,-1,-1,0,0} + 24\pi^2H_{0,0,-1,-1,0,-1} \notag\\
  &\quad - 34\pi^2H_{0,0,-1,-1,-1,0} - \tfrac{1030}{3}\pi^2H_{0,0,-1,-1,-1,-1} + \tfrac{1888}{9}\pi^2H_{0,-1,0,0,0,0} + 56\pi^2H_{0,-1,0,0,0,-1} \notag\\
  &\quad + 44\pi^2H_{0,-1,0,0,-1,0} - 36\pi^2H_{0,-1,0,0,-1,-1} + 48\pi^2H_{0,-1,0,-1,0,0} + 116\pi^2H_{0,-1,0,-1,0,-1} \notag\\
  &\quad + 2\pi^2H_{0,-1,0,-1,-1,0} - 510\pi^2H_{0,-1,0,-1,-1,-1} + \tfrac{1888}{9}\pi^2H_{0,-1,-1,0,0,0} - 64\pi^2H_{0,-1,-1,0,0,-1} \notag\\
  &\quad - 34\pi^2H_{0,-1,-1,0,-1,0} - 194\pi^2H_{0,-1,-1,0,-1,-1} + \tfrac{472}{9}\pi^2H_{0,-1,-1,-1,0,0} - 74\pi^2H_{0,-1,-1,-1,0,-1} \notag\\
  &\quad + \tfrac{25}{3}\pi^2H_{0,-1,-1,-1,-1,0} + \tfrac{25}{3}\pi^2H_{0,-1,-1,-1,-1,-1} - \tfrac{7424}{27}\pi^2H_{-1,0,0,0,0,0} + \tfrac{944}{3}\pi^2H_{-1,0,0,0,0,-1} \notag\\
  &\quad - \tfrac{212}{3}\pi^2H_{-1,0,0,0,-1,0} + \tfrac{1612}{3}\pi^2H_{-1,0,0,0,-1,-1} - \tfrac{28}{9}\pi^2H_{-1,0,0,-1,0,0} + 164\pi^2H_{-1,0,0,-1,0,-1} \notag\\
  &\quad + 136\pi^2H_{-1,0,0,-1,-1,0} - \tfrac{424}{3}\pi^2H_{-1,0,0,-1,-1,-1} + \tfrac{592}{9}\pi^2H_{-1,0,-1,0,0,0} + 168\pi^2H_{-1,0,-1,0,0,-1} \notag\\
  &\quad + 84\pi^2H_{-1,0,-1,0,-1,0} - 316\pi^2H_{-1,0,-1,0,-1,-1} + \tfrac{1564}{9}\pi^2H_{-1,0,-1,-1,0,0} - 240\pi^2H_{-1,0,-1,-1,0,-1} \notag\\
  &\quad + \tfrac{82}{3}\pi^2H_{-1,0,-1,-1,-1,0} + \tfrac{82}{3}\pi^2H_{-1,0,-1,-1,-1,-1} - \tfrac{1856}{27}\pi^2H_{-1,-1,0,0,0,0} + \tfrac{244}{3}\pi^2H_{-1,-1,0,0,0,-1} \notag\\
  &\quad - \tfrac{8}{3}\pi^2H_{-1,-1,0,0,-1,0} + \tfrac{200}{3}\pi^2H_{-1,-1,0,0,-1,-1} + \tfrac{148}{9}\pi^2H_{-1,-1,0,-1,0,0} - 16\pi^2H_{-1,-1,0,-1,0,-1} \notag\\
  &\quad + \tfrac{82}{3}\pi^2H_{-1,-1,0,-1,-1,0} + \tfrac{82}{3}\pi^2H_{-1,-1,0,-1,-1,-1} - \tfrac{464}{27}\pi^2H_{-1,-1,-1,0,0,0} + \tfrac{44}{3}\pi^2H_{-1,-1,-1,0,0,-1} \notag\\
  &\quad + \tfrac{2}{3}\pi^2H_{-1,-1,-1,0,-1,0} + \tfrac{82}{3}\pi^2H_{-1,-1,-1,0,-1,-1} - \tfrac{116}{27}\pi^2H_{-1,-1,-1,-1,0,0} + \tfrac{22}{3}\pi^2H_{-1,-1,-1,-1,0,-1} \notag\\
  &\quad - \tfrac{5}{9}\pi^2H_{-1,-1,-1,-1,-1,0} - \tfrac{5}{9}\pi^2H_{-1,-1,-1,-1,-1,-1} + \tfrac{160}{3}\zeta_3H_{0,0,0,0,-1} - 116\zeta_3H_{0,0,0,-1,0} \notag\\
  &\quad + \tfrac{1112}{3}\zeta_3H_{0,0,0,-1,-1} + \tfrac{1952}{9}\zeta_3H_{0,0,-1,0,0} + 144\zeta_3H_{0,0,-1,0,-1} - 300\zeta_3H_{0,0,-1,-1,0} \notag\\
  &\quad - \tfrac{2060}{3}\zeta_3H_{0,0,-1,-1,-1} + \tfrac{12208}{9}\zeta_3H_{0,-1,0,0,0} - 72\zeta_3H_{0,-1,0,0,-1} - 452\zeta_3H_{0,-1,0,-1,0} \notag\\
  &\quad - 1020\zeta_3H_{0,-1,0,-1,-1} + \tfrac{9916}{9}\zeta_3H_{0,-1,-1,0,0} - 388\zeta_3H_{0,-1,-1,0,-1} + 346\zeta_3H_{0,-1,-1,-1,0} \notag\\
  &\quad + \tfrac{50}{3}\zeta_3H_{0,-1,-1,-1,-1} - \tfrac{58688}{27}\zeta_3H_{-1,0,0,0,0} + \tfrac{3224}{3}\zeta_3H_{-1,0,0,0,-1} + 160\zeta_3H_{-1,0,0,-1,0} \notag\\
  &\quad - \tfrac{848}{3}\zeta_3H_{-1,0,0,-1,-1} + \tfrac{5152}{9}\zeta_3H_{-1,0,-1,0,0} - 632\zeta_3H_{-1,0,-1,0,-1} + 1124\zeta_3H_{-1,0,-1,-1,0} \notag\\
  &\quad + \tfrac{164}{3}\zeta_3H_{-1,0,-1,-1,-1} - \tfrac{14672}{27}\zeta_3H_{-1,-1,0,0,0} + \tfrac{400}{3}\zeta_3H_{-1,-1,0,0,-1} + 228\zeta_3H_{-1,-1,0,-1,0} \notag\\
  &\quad + \tfrac{164}{3}\zeta_3H_{-1,-1,0,-1,-1} - \tfrac{3668}{27}\zeta_3H_{-1,-1,-1,0,0} + \tfrac{164}{3}\zeta_3H_{-1,-1,-1,0,-1} - \tfrac{98}{3}\zeta_3H_{-1,-1,-1,-1,0} \notag\\
  &\quad - \tfrac{10}{9}\zeta_3H_{-1,-1,-1,-1,-1} + \tfrac{33}{5}\pi^4H_{0,0,0,-1} + \tfrac{34}{9}\pi^4H_{0,0,-1,0} - \tfrac{1567}{90}\pi^4H_{0,0,-1,-1} + \tfrac{1132}{135}\pi^4H_{0,-1,0,0} \notag\\
  &\quad - \tfrac{607}{30}\pi^4H_{0,-1,0,-1} + \tfrac{656}{45}\pi^4H_{0,-1,-1,0} + \tfrac{273}{20}\pi^4H_{0,-1,-1,-1} - \tfrac{17576}{405}\pi^4H_{-1,0,0,0} + \tfrac{44}{9}\pi^4H_{-1,0,0,-1} \notag\\
  &\quad - \tfrac{43}{5}\pi^4H_{-1,0,-1,0} + \tfrac{275}{6}\pi^4H_{-1,0,-1,-1} - \tfrac{4394}{405}\pi^4H_{-1,-1,0,0} + \tfrac{263}{30}\pi^4H_{-1,-1,0,-1} - \tfrac{259}{135}\pi^4H_{-1,-1,-1,0} \notag\\
  &\quad - \tfrac{23}{20}\pi^4H_{-1,-1,-1,-1} + \tfrac{674}{3}\zeta_3\pi^2H_{0,0,-1} - \tfrac{230}{9}\zeta_3\pi^2H_{0,-1,0} - 29\zeta_3\pi^2H_{0,-1,-1} - \tfrac{23476}{81}\zeta_3\pi^2H_{-1,0,0} \notag\\
  &\quad - \tfrac{574}{3}\zeta_3\pi^2H_{-1,0,-1} - \tfrac{2134}{27}\zeta_3\pi^2H_{-1,-1,0} - \tfrac{23}{3}\zeta_3\pi^2H_{-1,-1,-1} - \tfrac{2120}{3}\zeta_5H_{0,0,-1} + 2102\zeta_5H_{0,-1,0} \notag\\
  &\quad + \tfrac{2276}{3}\zeta_5H_{0,-1,-1} - \tfrac{17092}{9}\zeta_5H_{-1,0,0} + \tfrac{356}{3}\zeta_5H_{-1,0,-1} - \tfrac{1490}{3}\zeta_5H_{-1,-1,0} + \tfrac{344}{9}\zeta_5H_{-1,-1,-1} \notag\\
  &\quad + \tfrac{7135}{9}\zeta_3^2H_{0,-1} - \tfrac{6700}{9}\zeta_3^2H_{-1,0} - \tfrac{5135}{27}\zeta_3^2H_{-1,-1} + \tfrac{311}{45}\pi^6H_{0,-1} - \tfrac{12676}{2835}\pi^6H_{-1,0} - \tfrac{859}{405}\pi^6H_{-1,-1} \notag\\
  &\quad - \tfrac{11945}{486}\zeta_3\pi^4H_{0} - \tfrac{1843}{810}\zeta_3\pi^4H_{-1} - \tfrac{17183}{54}\zeta_5\pi^2H_{0} + \tfrac{3481}{27}\zeta_5\pi^2H_{-1} - \tfrac{47735}{18}\zeta_7H_{0} + \tfrac{9851}{9}\zeta_7H_{-1} + 275\zeta_{3,5} \notag\\
  &\quad - \tfrac{122786}{45}\zeta_3\zeta_5 - \tfrac{46532}{243}\zeta_3^2\pi^2 - \tfrac{16709389}{8164800}\pi^8\,. \label{eq:B4}
\end{align}

\begin{align}
  c^{(5)}_{-10} &= \tfrac{16}{225}\,, \notag\\
  c^{(5)}_{-9} &= \tfrac{7}{80}H_{0}\,, \notag\\
  c^{(5)}_{-8} &= - \tfrac{103}{1440}\pi^2\,, \notag\\
  c^{(5)}_{-7} &= - \tfrac{35}{144}H_{-1,0,0} + \tfrac{91}{576}\pi^2H_{0} - \tfrac{35}{288}\pi^2H_{-1} - \tfrac{2821}{2160}\zeta_3\,, \notag\\
  c^{(5)}_{-6} &= \tfrac{7}{4}H_{0,-1,0,0} - \tfrac{175}{144}H_{-1,0,0,0} - \tfrac{35}{144}H_{-1,-1,0,0} + \tfrac{7}{8}\pi^2H_{0,-1} - \tfrac{35}{216}\pi^2H_{-1,0} - \tfrac{35}{288}\pi^2H_{-1,-1} - \tfrac{133}{48}\zeta_3H_{0} \notag\\
  &\quad - \tfrac{35}{144}\zeta_3H_{-1} - \tfrac{3593}{17280}\pi^4\,, \notag\\
  c^{(5)}_{-5} &= - \tfrac{7}{3}H_{0,0,-1,0,0} + \tfrac{35}{4}H_{0,-1,0,0,0} + \tfrac{49}{12}H_{0,-1,-1,0,0} - \tfrac{875}{144}H_{-1,0,0,0,0} + \tfrac{155}{36}H_{-1,0,-1,0,0} - \tfrac{175}{144}H_{-1,-1,0,0,0} \notag\\
  &\quad - \tfrac{35}{144}H_{-1,-1,-1,0,0} - \tfrac{7}{6}\pi^2H_{0,0,-1} + \tfrac{7}{9}\pi^2H_{0,-1,0} + \tfrac{49}{24}\pi^2H_{0,-1,-1} - \tfrac{2585}{1728}\pi^2H_{-1,0,0} + \tfrac{155}{72}\pi^2H_{-1,0,-1} \notag\\
  &\quad - \tfrac{35}{216}\pi^2H_{-1,-1,0} - \tfrac{35}{288}\pi^2H_{-1,-1,-1} + \tfrac{49}{12}\zeta_3H_{0,-1} - \tfrac{115}{12}\zeta_3H_{-1,0} - \tfrac{35}{144}\zeta_3H_{-1,-1} + \tfrac{7043}{23040}\pi^4H_{0} \notag\\
  &\quad - \tfrac{4279}{10368}\pi^4H_{-1} + \tfrac{37621}{8640}\zeta_3\pi^2 + \tfrac{40781}{3600}\zeta_5\,, \notag\\
  c^{(5)}_{-4} &= \tfrac{10}{3}H_{0,0,0,-1,0,0} - \tfrac{35}{3}H_{0,0,-1,0,0,0} - \tfrac{49}{3}H_{0,0,-1,-1,0,0} + \tfrac{175}{4}H_{0,-1,0,0,0,0} - \tfrac{50}{3}H_{0,-1,0,-1,0,0} \notag\\
  &\quad + \tfrac{245}{12}H_{0,-1,-1,0,0,0} + \tfrac{7}{4}H_{0,-1,-1,-1,0,0} - \tfrac{4375}{144}H_{-1,0,0,0,0,0} + \tfrac{35}{3}H_{-1,0,0,-1,0,0} + \tfrac{775}{36}H_{-1,0,-1,0,0,0} \notag\\
  &\quad + \tfrac{575}{36}H_{-1,0,-1,-1,0,0} - \tfrac{875}{144}H_{-1,-1,0,0,0,0} + \tfrac{155}{36}H_{-1,-1,0,-1,0,0} - \tfrac{175}{144}H_{-1,-1,-1,0,0,0} \notag\\
  &\quad - \tfrac{35}{144}H_{-1,-1,-1,-1,0,0} + \tfrac{5}{3}\pi^2H_{0,0,0,-1} + \tfrac{7}{9}\pi^2H_{0,0,-1,0} - \tfrac{49}{6}\pi^2H_{0,0,-1,-1} + \tfrac{1015}{144}\pi^2H_{0,-1,0,0} \notag\\
  &\quad - \tfrac{25}{3}\pi^2H_{0,-1,0,-1} + \tfrac{28}{9}\pi^2H_{0,-1,-1,0} + \tfrac{7}{8}\pi^2H_{0,-1,-1,-1} - \tfrac{12925}{1728}\pi^2H_{-1,0,0,0} + \tfrac{35}{6}\pi^2H_{-1,0,0,-1} \notag\\
  &\quad + \tfrac{25}{27}\pi^2H_{-1,0,-1,0} + \tfrac{575}{72}\pi^2H_{-1,0,-1,-1} - \tfrac{2585}{1728}\pi^2H_{-1,-1,0,0} + \tfrac{155}{72}\pi^2H_{-1,-1,0,-1} - \tfrac{35}{216}\pi^2H_{-1,-1,-1,0} \notag\\
  &\quad - \tfrac{35}{288}\pi^2H_{-1,-1,-1,-1} - \tfrac{49}{3}\zeta_3H_{0,0,-1} + 52\zeta_3H_{0,-1,0} + \tfrac{7}{4}\zeta_3H_{0,-1,-1} - \tfrac{21215}{432}\zeta_3H_{-1,0,0} + \tfrac{575}{36}\zeta_3H_{-1,0,-1} \notag\\
  &\quad - \tfrac{115}{12}\zeta_3H_{-1,-1,0} - \tfrac{35}{144}\zeta_3H_{-1,-1,-1} + \tfrac{9101}{4320}\pi^4H_{0,-1} - \tfrac{79}{96}\pi^4H_{-1,0} - \tfrac{4279}{10368}\pi^4H_{-1,-1} - \tfrac{12485}{1728}\zeta_3\pi^2H_{0} \notag\\
  &\quad - \tfrac{6535}{1728}\zeta_3\pi^2H_{-1} - \tfrac{4693}{80}\zeta_5H_{0} + \tfrac{1645}{144}\zeta_5H_{-1} + \tfrac{10747}{6480}\zeta_3^2 - \tfrac{32881027}{43545600}\pi^6\,, \notag\\
  c^{(5)}_{-3} &= \tfrac{8}{3}H_{0,0,0,0,-1,0,0} + \tfrac{50}{3}H_{0,0,0,-1,0,0,0} + \tfrac{100}{3}H_{0,0,0,-1,-1,0,0} - \tfrac{175}{3}H_{0,0,-1,0,0,0,0} + \tfrac{56}{3}H_{0,0,-1,0,-1,0,0} \notag\\
  &\quad - \tfrac{245}{3}H_{0,0,-1,-1,0,0,0} - \tfrac{211}{3}H_{0,0,-1,-1,-1,0,0} + \tfrac{875}{4}H_{0,-1,0,0,0,0,0} + \tfrac{10}{3}H_{0,-1,0,0,-1,0,0} \notag\\
  &\quad - \tfrac{250}{3}H_{0,-1,0,-1,0,0,0} - \tfrac{332}{3}H_{0,-1,0,-1,-1,0,0} + \tfrac{1225}{12}H_{0,-1,-1,0,0,0,0} - \tfrac{200}{3}H_{0,-1,-1,0,-1,0,0} \notag\\
  &\quad + \tfrac{35}{4}H_{0,-1,-1,-1,0,0,0} + \tfrac{49}{12}H_{0,-1,-1,-1,-1,0,0} - \tfrac{21875}{144}H_{-1,0,0,0,0,0,0} + 80H_{-1,0,0,0,-1,0,0} \notag\\
  &\quad + \tfrac{175}{3}H_{-1,0,0,-1,0,0,0} + \tfrac{145}{3}H_{-1,0,0,-1,-1,0,0} + \tfrac{3875}{36}H_{-1,0,-1,0,0,0,0} - \tfrac{122}{9}H_{-1,0,-1,0,-1,0,0} \notag\\
  &\quad + \tfrac{2875}{36}H_{-1,0,-1,-1,0,0,0} + \tfrac{155}{36}H_{-1,0,-1,-1,-1,0,0} - \tfrac{4375}{144}H_{-1,-1,0,0,0,0,0} + \tfrac{35}{3}H_{-1,-1,0,0,-1,0,0} \notag\\
  &\quad + \tfrac{775}{36}H_{-1,-1,0,-1,0,0,0} + \tfrac{575}{36}H_{-1,-1,0,-1,-1,0,0} - \tfrac{875}{144}H_{-1,-1,-1,0,0,0,0} + \tfrac{155}{36}H_{-1,-1,-1,0,-1,0,0} \notag\\
  &\quad - \tfrac{175}{144}H_{-1,-1,-1,-1,0,0,0} - \tfrac{35}{144}H_{-1,-1,-1,-1,-1,0,0} + \tfrac{4}{3}\pi^2H_{0,0,0,0,-1} - \tfrac{25}{9}\pi^2H_{0,0,0,-1,0} \notag\\
  &\quad + \tfrac{50}{3}\pi^2H_{0,0,0,-1,-1} + \tfrac{95}{36}\pi^2H_{0,0,-1,0,0} + \tfrac{28}{3}\pi^2H_{0,0,-1,0,-1} - \tfrac{17}{9}\pi^2H_{0,0,-1,-1,0} - \tfrac{211}{6}\pi^2H_{0,0,-1,-1,-1} \notag\\
  &\quad + \tfrac{5075}{144}\pi^2H_{0,-1,0,0,0} + \tfrac{5}{3}\pi^2H_{0,-1,0,0,-1} + \tfrac{41}{9}\pi^2H_{0,-1,0,-1,0} - \tfrac{166}{3}\pi^2H_{0,-1,0,-1,-1} + \tfrac{3515}{144}\pi^2H_{0,-1,-1,0,0} \notag\\
  &\quad - \tfrac{100}{3}\pi^2H_{0,-1,-1,0,-1} + \tfrac{7}{9}\pi^2H_{0,-1,-1,-1,0} + \tfrac{49}{24}\pi^2H_{0,-1,-1,-1,-1} - \tfrac{64625}{1728}\pi^2H_{-1,0,0,0,0} \notag\\
  &\quad + 40\pi^2H_{-1,0,0,0,-1} + \tfrac{5}{3}\pi^2H_{-1,0,0,-1,0} + \tfrac{145}{6}\pi^2H_{-1,0,0,-1,-1} + \tfrac{5081}{432}\pi^2H_{-1,0,-1,0,0} \notag\\
  &\quad - \tfrac{61}{9}\pi^2H_{-1,0,-1,0,-1} + \tfrac{340}{27}\pi^2H_{-1,0,-1,-1,0} + \tfrac{155}{72}\pi^2H_{-1,0,-1,-1,-1} - \tfrac{12925}{1728}\pi^2H_{-1,-1,0,0,0} \notag\\
  &\quad + \tfrac{35}{6}\pi^2H_{-1,-1,0,0,-1} + \tfrac{25}{27}\pi^2H_{-1,-1,0,-1,0} + \tfrac{575}{72}\pi^2H_{-1,-1,0,-1,-1} - \tfrac{2585}{1728}\pi^2H_{-1,-1,-1,0,0} \notag\\
  &\quad + \tfrac{155}{72}\pi^2H_{-1,-1,-1,0,-1} - \tfrac{35}{216}\pi^2H_{-1,-1,-1,-1,0} - \tfrac{35}{288}\pi^2H_{-1,-1,-1,-1,-1} + \tfrac{100}{3}\zeta_3H_{0,0,0,-1} \notag\\
  &\quad - \tfrac{146}{3}\zeta_3H_{0,0,-1,0} - \tfrac{211}{3}\zeta_3H_{0,0,-1,-1} + \tfrac{2455}{12}\zeta_3H_{0,-1,0,0} - \tfrac{332}{3}\zeta_3H_{0,-1,0,-1} + 138\zeta_3H_{0,-1,-1,0} \notag\\
  &\quad + \tfrac{49}{12}\zeta_3H_{0,-1,-1,-1} - \tfrac{106075}{432}\zeta_3H_{-1,0,0,0} + \tfrac{145}{3}\zeta_3H_{-1,0,0,-1} + \tfrac{308}{3}\zeta_3H_{-1,0,-1,0} + \tfrac{155}{36}\zeta_3H_{-1,0,-1,-1} \notag\\
  &\quad - \tfrac{21215}{432}\zeta_3H_{-1,-1,0,0} + \tfrac{575}{36}\zeta_3H_{-1,-1,0,-1} - \tfrac{115}{12}\zeta_3H_{-1,-1,-1,0} - \tfrac{35}{144}\zeta_3H_{-1,-1,-1,-1} - \tfrac{1519}{1080}\pi^4H_{0,0,-1} \notag\\
  &\quad + \tfrac{217}{135}\pi^4H_{0,-1,0} + \tfrac{26969}{4320}\pi^4H_{0,-1,-1} - \tfrac{81077}{13824}\pi^4H_{-1,0,0} + \tfrac{59939}{12960}\pi^4H_{-1,0,-1} - \tfrac{79}{96}\pi^4H_{-1,-1,0} \notag\\
  &\quad - \tfrac{4279}{10368}\pi^4H_{-1,-1,-1} - \tfrac{211}{16}\zeta_3\pi^2H_{0,-1} - \tfrac{54751}{1296}\zeta_3\pi^2H_{-1,0} - \tfrac{6535}{1728}\zeta_3\pi^2H_{-1,-1} + \tfrac{289}{12}\zeta_5H_{0,-1} \notag\\
  &\quad - \tfrac{2743}{12}\zeta_5H_{-1,0} + \tfrac{1645}{144}\zeta_5H_{-1,-1} + \tfrac{1115}{288}\zeta_3^2H_{0} - \tfrac{37667}{432}\zeta_3^2H_{-1} + \tfrac{220823}{645120}\pi^6H_{0} - \tfrac{3650821}{2903040}\pi^6H_{-1} \notag\\
  &\quad + \tfrac{13060013}{1036800}\zeta_3\pi^4 + \tfrac{8247}{64}\zeta_5\pi^2 + \tfrac{389609}{5040}\zeta_7\,, \notag\\
  c^{(5)}_{-2} &= - \tfrac{32}{3}H_{0,0,0,0,0,-1,0,0} + \tfrac{40}{3}H_{0,0,0,0,-1,0,0,0} + \tfrac{62}{3}H_{0,0,0,0,-1,-1,0,0} + \tfrac{250}{3}H_{0,0,0,-1,0,0,0,0} \notag\\
  &\quad - \tfrac{56}{3}H_{0,0,0,-1,0,-1,0,0} + \tfrac{500}{3}H_{0,0,0,-1,-1,0,0,0} + \tfrac{754}{3}H_{0,0,0,-1,-1,-1,0,0} - \tfrac{875}{3}H_{0,0,-1,0,0,0,0,0} \notag\\
  &\quad - \tfrac{100}{3}H_{0,0,-1,0,0,-1,0,0} + \tfrac{280}{3}H_{0,0,-1,0,-1,0,0,0} + \tfrac{398}{3}H_{0,0,-1,0,-1,-1,0,0} - \tfrac{1225}{3}H_{0,0,-1,-1,0,0,0,0} \notag\\
  &\quad + \tfrac{152}{3}H_{0,0,-1,-1,0,-1,0,0} - \tfrac{1055}{3}H_{0,0,-1,-1,-1,0,0,0} - \tfrac{409}{3}H_{0,0,-1,-1,-1,-1,0,0} + \tfrac{4375}{4}H_{0,-1,0,0,0,0,0,0} \notag\\
  &\quad - \tfrac{148}{3}H_{0,-1,0,0,0,-1,0,0} + \tfrac{50}{3}H_{0,-1,0,0,-1,0,0,0} - \tfrac{344}{3}H_{0,-1,0,0,-1,-1,0,0} - \tfrac{1250}{3}H_{0,-1,0,-1,0,0,0,0} \notag\\
  &\quad + 132H_{0,-1,0,-1,0,-1,0,0} - \tfrac{1660}{3}H_{0,-1,0,-1,-1,0,0,0} - \tfrac{1118}{3}H_{0,-1,0,-1,-1,-1,0,0} + \tfrac{6125}{12}H_{0,-1,-1,0,0,0,0,0} \notag\\
  &\quad - \tfrac{410}{3}H_{0,-1,-1,0,0,-1,0,0} - \tfrac{1000}{3}H_{0,-1,-1,0,-1,0,0,0} - \tfrac{398}{3}H_{0,-1,-1,0,-1,-1,0,0} + \tfrac{175}{4}H_{0,-1,-1,-1,0,0,0,0} \notag\\
  &\quad - \tfrac{50}{3}H_{0,-1,-1,-1,0,-1,0,0} + \tfrac{245}{12}H_{0,-1,-1,-1,-1,0,0,0} + \tfrac{7}{4}H_{0,-1,-1,-1,-1,-1,0,0} - \tfrac{109375}{144}H_{-1,0,0,0,0,0,0,0} \notag\\
  &\quad + 382H_{-1,0,0,0,0,-1,0,0} + 400H_{-1,0,0,0,-1,0,0,0} + 510H_{-1,0,0,0,-1,-1,0,0} + \tfrac{875}{3}H_{-1,0,0,-1,0,0,0,0} \notag\\
  &\quad + \tfrac{584}{3}H_{-1,0,0,-1,0,-1,0,0} + \tfrac{725}{3}H_{-1,0,0,-1,-1,0,0,0} - \tfrac{625}{3}H_{-1,0,0,-1,-1,-1,0,0} + \tfrac{19375}{36}H_{-1,0,-1,0,0,0,0,0} \notag\\
  &\quad + 178H_{-1,0,-1,0,0,-1,0,0} - \tfrac{610}{9}H_{-1,0,-1,0,-1,0,0,0} - \tfrac{2792}{9}H_{-1,0,-1,0,-1,-1,0,0} + \tfrac{14375}{36}H_{-1,0,-1,-1,0,0,0,0} \notag\\
  &\quad - \tfrac{2372}{9}H_{-1,0,-1,-1,0,-1,0,0} + \tfrac{775}{36}H_{-1,0,-1,-1,-1,0,0,0} + \tfrac{575}{36}H_{-1,0,-1,-1,-1,-1,0,0} \notag\\
  &\quad - \tfrac{21875}{144}H_{-1,-1,0,0,0,0,0,0} + 80H_{-1,-1,0,0,0,-1,0,0} + \tfrac{175}{3}H_{-1,-1,0,0,-1,0,0,0} + \tfrac{145}{3}H_{-1,-1,0,0,-1,-1,0,0} \notag\\
  &\quad + \tfrac{3875}{36}H_{-1,-1,0,-1,0,0,0,0} - \tfrac{122}{9}H_{-1,-1,0,-1,0,-1,0,0} + \tfrac{2875}{36}H_{-1,-1,0,-1,-1,0,0,0} \notag\\
  &\quad + \tfrac{155}{36}H_{-1,-1,0,-1,-1,-1,0,0} - \tfrac{4375}{144}H_{-1,-1,-1,0,0,0,0,0} + \tfrac{35}{3}H_{-1,-1,-1,0,0,-1,0,0} + \tfrac{775}{36}H_{-1,-1,-1,0,-1,0,0,0} \notag\\
  &\quad + \tfrac{575}{36}H_{-1,-1,-1,0,-1,-1,0,0} - \tfrac{875}{144}H_{-1,-1,-1,-1,0,0,0,0} + \tfrac{155}{36}H_{-1,-1,-1,-1,0,-1,0,0} \notag\\
  &\quad - \tfrac{175}{144}H_{-1,-1,-1,-1,-1,0,0,0} - \tfrac{35}{144}H_{-1,-1,-1,-1,-1,-1,0,0} - \tfrac{16}{3}\pi^2H_{0,0,0,0,0,-1} - \tfrac{11}{9}\pi^2H_{0,0,0,0,-1,0} \notag\\
  &\quad + \tfrac{31}{3}\pi^2H_{0,0,0,0,-1,-1} - \tfrac{211}{18}\pi^2H_{0,0,0,-1,0,0} - \tfrac{28}{3}\pi^2H_{0,0,0,-1,0,-1} - \tfrac{127}{9}\pi^2H_{0,0,0,-1,-1,0} \notag\\
  &\quad + \tfrac{377}{3}\pi^2H_{0,0,0,-1,-1,-1} + \tfrac{475}{36}\pi^2H_{0,0,-1,0,0,0} - \tfrac{50}{3}\pi^2H_{0,0,-1,0,0,-1} - \tfrac{59}{9}\pi^2H_{0,0,-1,0,-1,0} \notag\\
  &\quad + \tfrac{199}{3}\pi^2H_{0,0,-1,0,-1,-1} - \tfrac{655}{36}\pi^2H_{0,0,-1,-1,0,0} + \tfrac{76}{3}\pi^2H_{0,0,-1,-1,0,-1} - \tfrac{323}{9}\pi^2H_{0,0,-1,-1,-1,0} \notag\\
  &\quad - \tfrac{409}{6}\pi^2H_{0,0,-1,-1,-1,-1} + \tfrac{25375}{144}\pi^2H_{0,-1,0,0,0,0} - \tfrac{74}{3}\pi^2H_{0,-1,0,0,0,-1} + \tfrac{197}{9}\pi^2H_{0,-1,0,0,-1,0} \notag\\
  &\quad - \tfrac{172}{3}\pi^2H_{0,-1,0,0,-1,-1} + \tfrac{23}{18}\pi^2H_{0,-1,0,-1,0,0} + 66\pi^2H_{0,-1,0,-1,0,-1} - \tfrac{271}{9}\pi^2H_{0,-1,0,-1,-1,0} \notag\\
  &\quad - \tfrac{559}{3}\pi^2H_{0,-1,0,-1,-1,-1} + \tfrac{17575}{144}\pi^2H_{0,-1,-1,0,0,0} - \tfrac{205}{3}\pi^2H_{0,-1,-1,0,0,-1} - \tfrac{301}{9}\pi^2H_{0,-1,-1,0,-1,0} \notag\\
  &\quad - \tfrac{199}{3}\pi^2H_{0,-1,-1,0,-1,-1} + \tfrac{1015}{144}\pi^2H_{0,-1,-1,-1,0,0} - \tfrac{25}{3}\pi^2H_{0,-1,-1,-1,0,-1} + \tfrac{28}{9}\pi^2H_{0,-1,-1,-1,-1,0} \notag\\
  &\quad + \tfrac{7}{8}\pi^2H_{0,-1,-1,-1,-1,-1} - \tfrac{323125}{1728}\pi^2H_{-1,0,0,0,0,0} + 191\pi^2H_{-1,0,0,0,0,-1} - \tfrac{55}{3}\pi^2H_{-1,0,0,0,-1,0} \notag\\
  &\quad + 255\pi^2H_{-1,0,0,0,-1,-1} + \tfrac{541}{36}\pi^2H_{-1,0,0,-1,0,0} + \tfrac{292}{3}\pi^2H_{-1,0,0,-1,0,-1} + 75\pi^2H_{-1,0,0,-1,-1,0} \notag\\
  &\quad - \tfrac{625}{6}\pi^2H_{-1,0,0,-1,-1,-1} + \tfrac{25405}{432}\pi^2H_{-1,0,-1,0,0,0} + 89\pi^2H_{-1,0,-1,0,0,-1} + \tfrac{1091}{27}\pi^2H_{-1,0,-1,0,-1,0} \notag\\
  &\quad - \tfrac{1396}{9}\pi^2H_{-1,0,-1,0,-1,-1} + \tfrac{42581}{432}\pi^2H_{-1,0,-1,-1,0,0} - \tfrac{1186}{9}\pi^2H_{-1,0,-1,-1,0,-1} + \tfrac{25}{27}\pi^2H_{-1,0,-1,-1,-1,0} \notag\\
  &\quad + \tfrac{575}{72}\pi^2H_{-1,0,-1,-1,-1,-1} - \tfrac{64625}{1728}\pi^2H_{-1,-1,0,0,0,0} + 40\pi^2H_{-1,-1,0,0,0,-1} + \tfrac{5}{3}\pi^2H_{-1,-1,0,0,-1,0} \notag\\
  &\quad + \tfrac{145}{6}\pi^2H_{-1,-1,0,0,-1,-1} + \tfrac{5081}{432}\pi^2H_{-1,-1,0,-1,0,0} - \tfrac{61}{9}\pi^2H_{-1,-1,0,-1,0,-1} + \tfrac{340}{27}\pi^2H_{-1,-1,0,-1,-1,0} \notag\\
  &\quad + \tfrac{155}{72}\pi^2H_{-1,-1,0,-1,-1,-1} - \tfrac{12925}{1728}\pi^2H_{-1,-1,-1,0,0,0} + \tfrac{35}{6}\pi^2H_{-1,-1,-1,0,0,-1} + \tfrac{25}{27}\pi^2H_{-1,-1,-1,0,-1,0} \notag\\
  &\quad + \tfrac{575}{72}\pi^2H_{-1,-1,-1,0,-1,-1} - \tfrac{2585}{1728}\pi^2H_{-1,-1,-1,-1,0,0} + \tfrac{155}{72}\pi^2H_{-1,-1,-1,-1,0,-1} \notag\\
  &\quad - \tfrac{35}{216}\pi^2H_{-1,-1,-1,-1,-1,0} - \tfrac{35}{288}\pi^2H_{-1,-1,-1,-1,-1,-1} + \tfrac{62}{3}\zeta_3H_{0,0,0,0,-1} - \tfrac{142}{3}\zeta_3H_{0,0,0,-1,0} \notag\\
  &\quad + \tfrac{754}{3}\zeta_3H_{0,0,0,-1,-1} - \tfrac{565}{9}\zeta_3H_{0,0,-1,0,0} + \tfrac{398}{3}\zeta_3H_{0,0,-1,0,-1} - \tfrac{950}{3}\zeta_3H_{0,0,-1,-1,0} - \tfrac{409}{3}\zeta_3H_{0,0,-1,-1,-1} \notag\\
  &\quad + \tfrac{12275}{12}\zeta_3H_{0,-1,0,0,0} - \tfrac{344}{3}\zeta_3H_{0,-1,0,0,-1} - \tfrac{1334}{3}\zeta_3H_{0,-1,0,-1,0} - \tfrac{1118}{3}\zeta_3H_{0,-1,0,-1,-1} \notag\\
  &\quad + \tfrac{22105}{36}\zeta_3H_{0,-1,-1,0,0} - \tfrac{398}{3}\zeta_3H_{0,-1,-1,0,-1} + 52\zeta_3H_{0,-1,-1,-1,0} + \tfrac{7}{4}\zeta_3H_{0,-1,-1,-1,-1} \notag\\
  &\quad - \tfrac{530375}{432}\zeta_3H_{-1,0,0,0,0} + 510\zeta_3H_{-1,0,0,0,-1} + \tfrac{182}{3}\zeta_3H_{-1,0,0,-1,0} - \tfrac{625}{3}\zeta_3H_{-1,0,0,-1,-1} \notag\\
  &\quad + \tfrac{38219}{108}\zeta_3H_{-1,0,-1,0,0} - \tfrac{2792}{9}\zeta_3H_{-1,0,-1,0,-1} + \tfrac{1598}{3}\zeta_3H_{-1,0,-1,-1,0} + \tfrac{575}{36}\zeta_3H_{-1,0,-1,-1,-1} \notag\\
  &\quad - \tfrac{106075}{432}\zeta_3H_{-1,-1,0,0,0} + \tfrac{145}{3}\zeta_3H_{-1,-1,0,0,-1} + \tfrac{308}{3}\zeta_3H_{-1,-1,0,-1,0} + \tfrac{155}{36}\zeta_3H_{-1,-1,0,-1,-1} \notag\\
  &\quad - \tfrac{21215}{432}\zeta_3H_{-1,-1,-1,0,0} + \tfrac{575}{36}\zeta_3H_{-1,-1,-1,0,-1} - \tfrac{115}{12}\zeta_3H_{-1,-1,-1,-1,0} - \tfrac{35}{144}\zeta_3H_{-1,-1,-1,-1,-1} \notag\\
  &\quad + \tfrac{2771}{540}\pi^4H_{0,0,0,-1} + \tfrac{251}{90}\pi^4H_{0,0,-1,0} - \tfrac{2201}{216}\pi^4H_{0,0,-1,-1} + \tfrac{230701}{17280}\pi^4H_{0,-1,0,0} - \tfrac{7333}{540}\pi^4H_{0,-1,0,-1} \notag\\
  &\quad + \tfrac{6097}{540}\pi^4H_{0,-1,-1,0} + \tfrac{9101}{4320}\pi^4H_{0,-1,-1,-1} - \tfrac{405385}{13824}\pi^4H_{-1,0,0,0} + \tfrac{249}{40}\pi^4H_{-1,0,0,-1} - \tfrac{10}{3}\pi^4H_{-1,0,-1,0} \notag\\
  &\quad + \tfrac{327959}{12960}\pi^4H_{-1,0,-1,-1} - \tfrac{81077}{13824}\pi^4H_{-1,-1,0,0} + \tfrac{59939}{12960}\pi^4H_{-1,-1,0,-1} - \tfrac{79}{96}\pi^4H_{-1,-1,-1,0} \notag\\
  &\quad - \tfrac{4279}{10368}\pi^4H_{-1,-1,-1,-1} + \tfrac{1453}{12}\zeta_3\pi^2H_{0,0,-1} + \tfrac{1996}{27}\zeta_3\pi^2H_{0,-1,0} + \tfrac{4729}{144}\zeta_3\pi^2H_{0,-1,-1} \notag\\
  &\quad - \tfrac{1046957}{5184}\zeta_3\pi^2H_{-1,0,0} - \tfrac{27173}{432}\zeta_3\pi^2H_{-1,0,-1} - \tfrac{54751}{1296}\zeta_3\pi^2H_{-1,-1,0} - \tfrac{6535}{1728}\zeta_3\pi^2H_{-1,-1,-1} \notag\\
  &\quad - 206\zeta_5H_{0,0,-1} + \tfrac{3667}{3}\zeta_5H_{0,-1,0} + \tfrac{1447}{4}\zeta_5H_{0,-1,-1} - \tfrac{160315}{144}\zeta_5H_{-1,0,0} + \tfrac{4415}{36}\zeta_5H_{-1,0,-1} \notag\\
  &\quad - \tfrac{2743}{12}\zeta_5H_{-1,-1,0} + \tfrac{1645}{144}\zeta_5H_{-1,-1,-1} + \tfrac{25297}{36}\zeta_3^2H_{0,-1} - \tfrac{17371}{36}\zeta_3^2H_{-1,0} - \tfrac{37667}{432}\zeta_3^2H_{-1,-1} \notag\\
  &\quad + \tfrac{4021657}{725760}\pi^6H_{0,-1} - \tfrac{707981}{241920}\pi^6H_{-1,0} - \tfrac{3650821}{2903040}\pi^6H_{-1,-1} - \tfrac{245945}{13824}\zeta_3\pi^4H_{0} - \tfrac{3029357}{622080}\zeta_3\pi^4H_{-1} \notag\\
  &\quad - \tfrac{118609}{576}\zeta_5\pi^2H_{0} + \tfrac{17143}{576}\zeta_5\pi^2H_{-1} - \tfrac{28571}{16}\zeta_7H_{0} + \tfrac{75955}{144}\zeta_7H_{-1} + \tfrac{21694}{75}\zeta_{3,5} - \tfrac{479761}{270}\zeta_3\zeta_5 - \tfrac{751033}{6480}\zeta_3^2\pi^2 \notag\\
  &\quad - \tfrac{26402286431}{13063680000}\pi^8\,. \label{eq:B5}
\end{align}

\subsection{Non-planar boxes}

\noindent As for the planar boxes, we do not show all calculated orders here. The complete results can be found in the ancillary files. In the normalisation of Eq.~\eqref{eq:np-box}, with the harmonic polylogarithms $H_w(x)$, $H_w(y)$ and $H_w(x/y)$ defined above, we have
\allowdisplaybreaks
\begin{align}
  xy\,d^{(2)}_{-4} &= \tfrac{5}{2} + \tfrac{1}{2} x + \tfrac{1}{2} y\,, \notag\\
  xy\,d^{(2)}_{-3} &= - 3H_{0}(x) - 3H_{0}(y) + \tfrac{3}{2} + x\Big[\tfrac{1}{2}H_{0}(x) - \tfrac{1}{2}H_{0}(y) + \tfrac{3}{2}\Big] + y\Big[- \tfrac{1}{2}H_{0}(x) + \tfrac{1}{2}H_{0}(y) + \tfrac{3}{2}\Big]\,, \notag\\
  xy\,d^{(2)}_{-2} &= 2H_{0,0}(x) + 4H_{0}(x)H_{0}(y) + 2H_{0,0}(y) - \tfrac{7}{4}\pi^2 - 3H_{0}(x) - 3H_{0}(y) - 6 + x\Big[- \tfrac{1}{2}\pi^2 + 3H_{0}(x) \notag\\
  &\quad - 3H_{0}(y) - 6\Big] + y\Big[- \tfrac{1}{2}\pi^2 - 3H_{0}(x) + 3H_{0}(y) - 6\Big]\,, \notag\\
  xy\,d^{(2)}_{-1} &= 4H_{0,0,0}(x) - 4H_{0,0}(x)H_{0}(y) - 2H_{0}(x)H_{0,0}(y) - 2H_{-1,0,0}(x/y) + 2H_{0,0,0}(y) + \tfrac{9}{2}\pi^2H_{0}(x) \notag\\
  &\quad - \pi^2H_{-1}(x/y) + \tfrac{7}{2}\pi^2H_{0}(y) - \tfrac{74}{3}\zeta_3 + 6H_{0,0}(x) + 6H_{0}(x)H_{0}(y) + 6H_{0,0}(y) + \tfrac{11}{4}\pi^2 + 12H_{0}(x) \notag\\
  &\quad + 12H_{0}(y) + 24 + x\Big[- 2H_{0,0,0}(x) + 2H_{0,0}(x)H_{0}(y) - 2H_{0}(x)H_{0,0}(y) + 2H_{0,0,0}(y) - \tfrac{7}{4}\pi^2H_{0}(x) \notag\\
  &\quad + \tfrac{7}{4}\pi^2H_{0}(y) - \tfrac{83}{6}\zeta_3 + 6H_{0,0}(x) - 6H_{0}(x)H_{0}(y) + 6H_{0,0}(y) + \tfrac{11}{4}\pi^2 - 12H_{0}(x) + 12H_{0}(y) + 24\Big] \notag\\
  &\quad + y\Big[2H_{0,0,0}(x) - 2H_{0,0}(x)H_{0}(y) + 2H_{0}(x)H_{0,0}(y) - 2H_{0,0,0}(y) + \tfrac{7}{4}\pi^2H_{0}(x) - \tfrac{7}{4}\pi^2H_{0}(y) - \tfrac{83}{6}\zeta_3 \notag\\
  &\quad + 6H_{0,0}(x) - 6H_{0}(x)H_{0}(y) + 6H_{0,0}(y) + \tfrac{11}{4}\pi^2 + 12H_{0}(x) - 12H_{0}(y) + 24\Big]\,, \notag\\
  xy\,d^{(2)}_{0} &= - 24H_{0,0,0,0}(x) + 18H_{0,-1,0,0}(x) - 6H_{-1,-1,0,0}(x) + 6H_{-1,0,0}(x)H_{0}(y) - 6H_{0}(x)H_{0,0,0}(y) \notag\\
  &\quad + 6H_{0}(x)H_{-1,0,0}(y) + 2H_{0,-1,0,0}(x/y) + 8H_{-1,0,0,0}(x/y) + 2H_{-1,-1,0,0}(x/y) \notag\\
  &\quad + 6H_{-1,0,0}(x/y)H_{0}(y) - 12H_{0,0,0,0}(y) + 18H_{0,-1,0,0}(y) - 6H_{-1,-1,0,0}(y) - 11\pi^2H_{0,0}(x) \notag\\
  &\quad + 9\pi^2H_{0,-1}(x) + \pi^2H_{-1,0}(x) - 3\pi^2H_{-1,-1}(x) - 8\pi^2H_{0}(x)H_{0}(y) + 3\pi^2H_{0}(x)H_{-1}(y) \notag\\
  &\quad + 3\pi^2H_{-1}(x)H_{0}(y) + \pi^2H_{0,-1}(x/y) + \pi^2H_{-1,0}(x/y) + \pi^2H_{-1,-1}(x/y) + 3\pi^2H_{-1}(x/y)H_{0}(y) \notag\\
  &\quad - 8\pi^2H_{0,0}(y) + 9\pi^2H_{0,-1}(y) + \pi^2H_{-1,0}(y) - 3\pi^2H_{-1,-1}(y) + 40\zeta_3H_{0}(x) - 6\zeta_3H_{-1}(x) \notag\\
  &\quad + 2\zeta_3H_{-1}(x/y) + 42\zeta_3H_{0}(y) - 6\zeta_3H_{-1}(y) - \tfrac{863}{240}\pi^4 - 12H_{0,0,0}(x) + 24H_{-1,0,0}(x) \notag\\
  &\quad - 12H_{0,0}(x)H_{0}(y) - 12H_{0}(x)H_{0,0}(y) - 12H_{0,0,0}(y) + 24H_{-1,0,0}(y) - \tfrac{11}{2}\pi^2H_{0}(x) + 12\pi^2H_{-1}(x) \notag\\
  &\quad - \tfrac{11}{2}\pi^2H_{0}(y) + 12\pi^2H_{-1}(y) + 8\zeta_3 - 24H_{0,0}(x) - 24H_{0}(x)H_{0}(y) - 24H_{0,0}(y) - 11\pi^2 - 48H_{0}(x) \notag\\
  &\quad - 48H_{0}(y) - 96 + x\Big[8H_{0,0,0,0}(x) - 6H_{-1,-1,0,0}(x) - 8H_{0,0,0}(x)H_{0}(y) + 6H_{-1,0,0}(x)H_{0}(y) \notag\\
  &\quad + 2H_{0,0}(x)H_{0,0}(y) - 2H_{0}(x)H_{0,0,0}(y) - 10H_{0,-1,0,0}(x/y) + 2H_{0,0,0,0}(y) + 2\pi^2H_{0,0}(x) \notag\\
  &\quad + \pi^2H_{-1,0}(x) - 3\pi^2H_{-1,-1}(x) - 3\pi^2H_{0}(x)H_{0}(y) + 3\pi^2H_{-1}(x)H_{0}(y) - 5\pi^2H_{0,-1}(x/y) \notag\\
  &\quad - \tfrac{58}{3}\zeta_3H_{0}(x) - 6\zeta_3H_{-1}(x) + \tfrac{76}{3}\zeta_3H_{0}(y) - \tfrac{91}{240}\pi^4 - 36H_{0,0,0}(x) + 24H_{-1,0,0}(x) + 12H_{0,0}(x)H_{0}(y) \notag\\
  &\quad - 12H_{0}(x)H_{0,0}(y) + 24H_{-1,0,0}(x/y) + 12H_{0,0,0}(y) - \tfrac{37}{2}\pi^2H_{0}(x) + 12\pi^2H_{-1}(x) + 12\pi^2H_{-1}(x/y) \notag\\
  &\quad + \tfrac{13}{2}\pi^2H_{0}(y) + 8\zeta_3 - 24H_{0,0}(x) + 24H_{0}(x)H_{0}(y) - 24H_{0,0}(y) - 11\pi^2 + 48H_{0}(x) - 48H_{0}(y) - 96\Big] \notag\\
  &\quad + y\Big[- 8H_{0,0,0,0}(x) + 8H_{0,0,0}(x)H_{0}(y) - 8H_{0,0}(x)H_{0,0}(y) + 2H_{0}(x)H_{0,0,0}(y) + 6H_{0}(x)H_{-1,0,0}(y) \notag\\
  &\quad + 10H_{0,-1,0,0}(x/y) - 2H_{0,0,0,0}(y) - 6H_{-1,-1,0,0}(y) - 5\pi^2H_{0,0}(x) + 2\pi^2H_{0}(x)H_{0}(y) \notag\\
  &\quad + 3\pi^2H_{0}(x)H_{-1}(y) + 5\pi^2H_{0,-1}(x/y) - 3\pi^2H_{0,0}(y) + \pi^2H_{-1,0}(y) - 3\pi^2H_{-1,-1}(y) + \tfrac{76}{3}\zeta_3H_{0}(x) \notag\\
  &\quad - \tfrac{58}{3}\zeta_3H_{0}(y) - 6\zeta_3H_{-1}(y) - \tfrac{431}{240}\pi^4 - 12H_{0,0,0}(x) + 12H_{0,0}(x)H_{0}(y) - 12H_{0}(x)H_{0,0}(y) \notag\\
  &\quad + 24H_{-1,0,0}(x/y) - 12H_{0,0,0}(y) + 24H_{-1,0,0}(y) - \tfrac{11}{2}\pi^2H_{0}(x) + 12\pi^2H_{-1}(x/y) - \tfrac{13}{2}\pi^2H_{0}(y) \notag\\
  &\quad + 12\pi^2H_{-1}(y) + 8\zeta_3 - 24H_{0,0}(x) + 24H_{0}(x)H_{0}(y) - 24H_{0,0}(y) - 11\pi^2 - 48H_{0}(x) + 48H_{0}(y) \notag\\
  &\quad - 96\Big]\,. \label{eq:D2}
\end{align}

\begin{align}
  xy\,d^{(3)}_{-6} &= \tfrac{22}{45} + \tfrac{11}{15} x + \tfrac{11}{15} y\,, \notag\\
  xy\,d^{(3)}_{-5} &= - \tfrac{29}{30}H_{0}(x) - \tfrac{29}{30}H_{0}(y) + \tfrac{7}{15} + x\Big[\tfrac{7}{15}H_{0}(x) - \tfrac{41}{30}H_{0}(y) + \tfrac{13}{9}\Big] + y\Big[- \tfrac{41}{30}H_{0}(x) + \tfrac{7}{15}H_{0}(y) + \tfrac{13}{9}\Big] \notag\\
  &\quad + \tfrac{1}{5} xy\,, \notag\\
  xy\,d^{(3)}_{-4} &= \tfrac{7}{5}H_{0,0}(x) + \tfrac{19}{10}H_{0}(x)H_{0}(y) + \tfrac{7}{5}H_{0,0}(y) - \tfrac{13}{36}\pi^2 - \tfrac{7}{5}H_{0}(x) - \tfrac{7}{5}H_{0}(y) - \tfrac{8}{5} + x\Big[- \tfrac{2}{5}H_{0,0}(x) \notag\\
  &\quad - \tfrac{2}{5}H_{0}(x)H_{0}(y) + \tfrac{8}{5}H_{0,0}(y) - \tfrac{19}{18}\pi^2 + \tfrac{7}{3}H_{0}(x) - \tfrac{13}{3}H_{0}(y) - \tfrac{308}{45}\Big] + y\Big[\tfrac{8}{5}H_{0,0}(x) - \tfrac{2}{5}H_{0}(x)H_{0}(y) \notag\\
  &\quad - \tfrac{2}{5}H_{0,0}(y) - \tfrac{19}{18}\pi^2 - \tfrac{13}{3}H_{0}(x) + \tfrac{7}{3}H_{0}(y) - \tfrac{308}{45}\Big] + xy\Big[\tfrac{2}{5}H_{0}(x) + \tfrac{2}{5}H_{0}(y) - \tfrac{4}{5}\Big]\,, \notag\\
  xy\,d^{(3)}_{-3} &= \tfrac{3}{10}H_{0,0,0}(x) - \tfrac{27}{10}H_{0,0}(x)H_{0}(y) - \tfrac{11}{5}H_{0}(x)H_{0,0}(y) - \tfrac{1}{2}H_{-1,0,0}(x/y) - \tfrac{1}{5}H_{0,0,0}(y) + \tfrac{47}{24}\pi^2H_{0}(x) \notag\\
  &\quad - \tfrac{1}{4}\pi^2H_{-1}(x/y) + \tfrac{41}{24}\pi^2H_{0}(y) - \tfrac{601}{90}\zeta_3 + \tfrac{21}{5}H_{0,0}(x) + \tfrac{21}{5}H_{0}(x)H_{0}(y) + \tfrac{21}{5}H_{0,0}(y) + \tfrac{19}{12}\pi^2 \notag\\
  &\quad + \tfrac{24}{5}H_{0}(x) + \tfrac{24}{5}H_{0}(y) + \tfrac{56}{15} + x\Big[- \tfrac{9}{5}H_{0,0,0}(x) + \tfrac{11}{5}H_{0,0}(x)H_{0}(y) - \tfrac{9}{5}H_{0}(x)H_{0,0}(y) + \tfrac{27}{10}H_{0,0,0}(y) \notag\\
  &\quad - \tfrac{1}{2}H_{-1,0,0}(y) - \tfrac{25}{12}\pi^2H_{0}(x) + \tfrac{33}{8}\pi^2H_{0}(y) - \tfrac{1}{4}\pi^2H_{-1}(y) - \tfrac{613}{30}\zeta_3 + 3H_{0,0}(x) - 7H_{0}(x)H_{0}(y) \notag\\
  &\quad + 13H_{0,0}(y) + \tfrac{97}{36}\pi^2 - \tfrac{152}{15}H_{0}(x) + \tfrac{308}{15}H_{0}(y) + \tfrac{1412}{45}\Big] + y\Big[\tfrac{27}{10}H_{0,0,0}(x) - \tfrac{1}{2}H_{-1,0,0}(x) \notag\\
  &\quad - \tfrac{9}{5}H_{0,0}(x)H_{0}(y) + \tfrac{11}{5}H_{0}(x)H_{0,0}(y) - \tfrac{9}{5}H_{0,0,0}(y) + \tfrac{33}{8}\pi^2H_{0}(x) - \tfrac{1}{4}\pi^2H_{-1}(x) - \tfrac{25}{12}\pi^2H_{0}(y) \notag\\
  &\quad - \tfrac{613}{30}\zeta_3 + 13H_{0,0}(x) - 7H_{0}(x)H_{0}(y) + 3H_{0,0}(y) + \tfrac{97}{36}\pi^2 + \tfrac{308}{15}H_{0}(x) - \tfrac{152}{15}H_{0}(y) + \tfrac{1412}{45}\Big] \notag\\
  &\quad + xy\Big[\tfrac{4}{5}H_{0,0}(x) + \tfrac{4}{5}H_{0}(x)H_{0}(y) + \tfrac{4}{5}H_{0,0}(y) + \tfrac{17}{12}\pi^2 - \tfrac{8}{5}H_{0}(x) - \tfrac{8}{5}H_{0}(y) + \tfrac{16}{5}\Big]\,, \notag\\
  xy\,d^{(3)}_{-2} &= - \tfrac{72}{5}H_{0,0,0,0}(x) + 12H_{0,-1,0,0}(x) - 6H_{-1,-1,0,0}(x) - \tfrac{9}{10}H_{0,0,0}(x)H_{0}(y) + 6H_{-1,0,0}(x)H_{0}(y) \notag\\
  &\quad + \tfrac{3}{5}H_{0,0}(x)H_{0,0}(y) - \tfrac{17}{5}H_{0}(x)H_{0,0,0}(y) + 6H_{0}(x)H_{-1,0,0}(y) + \tfrac{1}{2}H_{0,-1,0,0}(x/y) + 3H_{-1,0,0,0}(x/y) \notag\\
  &\quad + \tfrac{3}{2}H_{-1,-1,0,0}(x/y) + \tfrac{5}{2}H_{-1,0,0}(x/y)H_{0}(y) - \tfrac{47}{5}H_{0,0,0,0}(y) + 12H_{0,-1,0,0}(y) - 6H_{-1,-1,0,0}(y) \notag\\
  &\quad - \tfrac{17}{2}\pi^2H_{0,0}(x) + 6\pi^2H_{0,-1}(x) + \pi^2H_{-1,0}(x) - 3\pi^2H_{-1,-1}(x) - \tfrac{55}{8}\pi^2H_{0}(x)H_{0}(y) \notag\\
  &\quad + 3\pi^2H_{0}(x)H_{-1}(y) + 3\pi^2H_{-1}(x)H_{0}(y) + \tfrac{1}{4}\pi^2H_{0,-1}(x/y) + \tfrac{1}{4}\pi^2H_{-1,0}(x/y) + \tfrac{3}{4}\pi^2H_{-1,-1}(x/y) \notag\\
  &\quad + \tfrac{5}{4}\pi^2H_{-1}(x/y)H_{0}(y) - \tfrac{29}{4}\pi^2H_{0,0}(y) + 6\pi^2H_{0,-1}(y) + \pi^2H_{-1,0}(y) - 3\pi^2H_{-1,-1}(y) + \tfrac{263}{15}\zeta_3H_{0}(x) \notag\\
  &\quad - 6\zeta_3H_{-1}(x) + \tfrac{3}{2}\zeta_3H_{-1}(x/y) + \tfrac{571}{30}\zeta_3H_{0}(y) - 6\zeta_3H_{-1}(y) - \tfrac{7063}{2700}\pi^4 - \tfrac{63}{5}H_{0,0,0}(x) + 18H_{-1,0,0}(x) \notag\\
  &\quad - \tfrac{63}{5}H_{0,0}(x)H_{0}(y) - \tfrac{63}{5}H_{0}(x)H_{0,0}(y) - \tfrac{63}{5}H_{0,0,0}(y) + 18H_{-1,0,0}(y) - \tfrac{19}{4}\pi^2H_{0}(x) + 9\pi^2H_{-1}(x) \notag\\
  &\quad - \tfrac{19}{4}\pi^2H_{0}(y) + 9\pi^2H_{-1}(y) - \tfrac{113}{15}\zeta_3 - \tfrac{72}{5}H_{0,0}(x) - \tfrac{72}{5}H_{0}(x)H_{0}(y) - \tfrac{72}{5}H_{0,0}(y) - \tfrac{14}{3}\pi^2 - \tfrac{56}{5}H_{0}(x) \notag\\
  &\quad - \tfrac{56}{5}H_{0}(y) + \tfrac{32}{5} + x\Big[\tfrac{72}{5}H_{0,0,0,0}(x) - 6H_{-1,-1,0,0}(x) - \tfrac{48}{5}H_{0,0,0}(x)H_{0}(y) + 6H_{-1,0,0}(x)H_{0}(y) \notag\\
  &\quad + \tfrac{2}{5}H_{0,0}(x)H_{0,0}(y) + \tfrac{22}{5}H_{0}(x)H_{0,0,0}(y) - 10H_{0,-1,0,0}(x/y) - \tfrac{103}{5}H_{0,0,0,0}(y) + \tfrac{21}{2}H_{0,-1,0,0}(y) \notag\\
  &\quad + 3H_{-1,0,0,0}(y) - \tfrac{1}{2}H_{-1,-1,0,0}(y) + \tfrac{9}{2}\pi^2H_{0,0}(x) + \pi^2H_{-1,0}(x) - 3\pi^2H_{-1,-1}(x) - \tfrac{1}{2}\pi^2H_{0}(x)H_{0}(y) \notag\\
  &\quad + 3\pi^2H_{-1}(x)H_{0}(y) - 5\pi^2H_{0,-1}(x/y) - \tfrac{41}{4}\pi^2H_{0,0}(y) + \tfrac{21}{4}\pi^2H_{0,-1}(y) + \tfrac{7}{12}\pi^2H_{-1,0}(y) \notag\\
  &\quad - \tfrac{1}{4}\pi^2H_{-1,-1}(y) - \tfrac{293}{15}\zeta_3H_{0}(x) - 6\zeta_3H_{-1}(x) + \tfrac{797}{15}\zeta_3H_{0}(y) - \tfrac{1}{2}\zeta_3H_{-1}(y) - \tfrac{20711}{21600}\pi^4 - 39H_{0,0,0}(x) \notag\\
  &\quad + 18H_{-1,0,0}(x) + 13H_{0,0}(x)H_{0}(y) - H_{0}(x)H_{0,0}(y) + 22H_{-1,0,0}(x/y) - 17H_{0,0,0}(y) + 8H_{-1,0,0}(y) \notag\\
  &\quad - \tfrac{209}{12}\pi^2H_{0}(x) + 9\pi^2H_{-1}(x) + 11\pi^2H_{-1}(x/y) + \tfrac{35}{12}\pi^2H_{0}(y) + 4\pi^2H_{-1}(y) + \tfrac{133}{9}\zeta_3 - \tfrac{48}{5}H_{0,0}(x) \notag\\
  &\quad + \tfrac{152}{5}H_{0}(x)H_{0}(y) - \tfrac{308}{5}H_{0,0}(y) - \tfrac{91}{9}\pi^2 + \tfrac{608}{15}H_{0}(x) - \tfrac{1412}{15}H_{0}(y) - \tfrac{6188}{45}\Big] + y\Big[- \tfrac{153}{5}H_{0,0,0,0}(x) \notag\\
  &\quad + \tfrac{21}{2}H_{0,-1,0,0}(x) + 3H_{-1,0,0,0}(x) - \tfrac{1}{2}H_{-1,-1,0,0}(x) + \tfrac{72}{5}H_{0,0,0}(x)H_{0}(y) - \tfrac{48}{5}H_{0,0}(x)H_{0,0}(y) \notag\\
  &\quad + \tfrac{2}{5}H_{0}(x)H_{0,0,0}(y) + 6H_{0}(x)H_{-1,0,0}(y) + 10H_{0,-1,0,0}(x/y) + \tfrac{22}{5}H_{0,0,0,0}(y) - 6H_{-1,-1,0,0}(y) \notag\\
  &\quad - \tfrac{61}{4}\pi^2H_{0,0}(x) + \tfrac{21}{4}\pi^2H_{0,-1}(x) + \tfrac{7}{12}\pi^2H_{-1,0}(x) - \tfrac{1}{4}\pi^2H_{-1,-1}(x) + \tfrac{9}{2}\pi^2H_{0}(x)H_{0}(y) \notag\\
  &\quad + 3\pi^2H_{0}(x)H_{-1}(y) + 5\pi^2H_{0,-1}(x/y) - \tfrac{1}{2}\pi^2H_{0,0}(y) + \pi^2H_{-1,0}(y) - 3\pi^2H_{-1,-1}(y) + \tfrac{797}{15}\zeta_3H_{0}(x) \notag\\
  &\quad - \tfrac{1}{2}\zeta_3H_{-1}(x) - \tfrac{293}{15}\zeta_3H_{0}(y) - 6\zeta_3H_{-1}(y) - \tfrac{51311}{21600}\pi^4 - 39H_{0,0,0}(x) + 8H_{-1,0,0}(x) \notag\\
  &\quad + 21H_{0,0}(x)H_{0}(y) - 9H_{0}(x)H_{0,0}(y) + 22H_{-1,0,0}(x/y) - 17H_{0,0,0}(y) + 18H_{-1,0,0}(y) - \tfrac{97}{12}\pi^2H_{0}(x) \notag\\
  &\quad + 4\pi^2H_{-1}(x) + 11\pi^2H_{-1}(x/y) - \tfrac{77}{12}\pi^2H_{0}(y) + 9\pi^2H_{-1}(y) + \tfrac{133}{9}\zeta_3 - \tfrac{308}{5}H_{0,0}(x) \notag\\
  &\quad + \tfrac{152}{5}H_{0}(x)H_{0}(y) - \tfrac{48}{5}H_{0,0}(y) - \tfrac{91}{9}\pi^2 - \tfrac{1412}{15}H_{0}(x) + \tfrac{608}{15}H_{0}(y) - \tfrac{6188}{45}\Big] + xy\Big[- \tfrac{42}{5}H_{0,0,0}(x) \notag\\
  &\quad + 10H_{-1,0,0}(x) + \tfrac{8}{5}H_{0,0}(x)H_{0}(y) + \tfrac{8}{5}H_{0}(x)H_{0,0}(y) - \tfrac{42}{5}H_{0,0,0}(y) + 10H_{-1,0,0}(y) - \tfrac{13}{6}\pi^2H_{0}(x) \notag\\
  &\quad + 5\pi^2H_{-1}(x) - \tfrac{13}{6}\pi^2H_{0}(y) + 5\pi^2H_{-1}(y) - \tfrac{29}{5}\zeta_3 - \tfrac{16}{5}H_{0,0}(x) - \tfrac{16}{5}H_{0}(x)H_{0}(y) - \tfrac{16}{5}H_{0,0}(y) \notag\\
  &\quad - \tfrac{17}{3}\pi^2 + \tfrac{32}{5}H_{0}(x) + \tfrac{32}{5}H_{0}(y) - \tfrac{64}{5}\Big]\,. \label{eq:D3}
\end{align}

\begin{align}
  xy\,d^{(4)}_{-8} &= \tfrac{31}{672} + \tfrac{599}{2016} x + \tfrac{599}{2016} y\,, \notag\\
  xy\,d^{(4)}_{-7} &= - \tfrac{65}{504}H_{0}(x) - \tfrac{65}{504}H_{0}(y) + \tfrac{37}{672} + x\Big[\tfrac{37}{336}H_{0}(x) - \tfrac{827}{1008}H_{0}(y) + \tfrac{13}{32}\Big] + y\Big[- \tfrac{827}{1008}H_{0}(x) + \tfrac{37}{336}H_{0}(y) \notag\\
  &\quad + \tfrac{13}{32}\Big] + \tfrac{1}{12} xy\,, \notag\\
  xy\,d^{(4)}_{-6} &= \tfrac{37}{126}H_{0,0}(x) + \tfrac{22}{63}H_{0}(x)H_{0}(y) + \tfrac{37}{126}H_{0,0}(y) - \tfrac{5}{288}\pi^2 - \tfrac{37}{168}H_{0}(x) - \tfrac{37}{168}H_{0}(y) - \tfrac{15}{112} \notag\\
  &\quad + x\Big[- \tfrac{4}{21}H_{0,0}(x) - \tfrac{4}{21}H_{0}(x)H_{0}(y) + \tfrac{38}{21}H_{0,0}(y) - \tfrac{205}{432}\pi^2 + \tfrac{37}{72}H_{0}(x) - \tfrac{13}{8}H_{0}(y) - \tfrac{2339}{1008}\Big] \notag\\
  &\quad + y\Big[\tfrac{38}{21}H_{0,0}(x) - \tfrac{4}{21}H_{0}(x)H_{0}(y) - \tfrac{4}{21}H_{0,0}(y) - \tfrac{205}{432}\pi^2 - \tfrac{13}{8}H_{0}(x) + \tfrac{37}{72}H_{0}(y) - \tfrac{2339}{1008}\Big] + xy\Big[\tfrac{1}{6}H_{0}(x) \notag\\
  &\quad + \tfrac{1}{6}H_{0}(y) - \tfrac{131}{252}\Big]\,, \notag\\
  xy\,d^{(4)}_{-5} &= - \tfrac{2}{7}H_{0,0,0}(x) - \tfrac{46}{63}H_{0,0}(x)H_{0}(y) - \tfrac{85}{126}H_{0}(x)H_{0,0}(y) - \tfrac{1}{18}H_{-1,0,0}(x/y) - \tfrac{43}{126}H_{0,0,0}(y) \notag\\
  &\quad + \tfrac{59}{216}\pi^2H_{0}(x) - \tfrac{1}{36}\pi^2H_{-1}(x/y) + \tfrac{53}{216}\pi^2H_{0}(y) - \tfrac{11}{21}\zeta_3 + \tfrac{37}{42}H_{0,0}(x) + \tfrac{37}{42}H_{0}(x)H_{0}(y) + \tfrac{37}{42}H_{0,0}(y) \notag\\
  &\quad + \tfrac{89}{288}\pi^2 + \tfrac{15}{28}H_{0}(x) + \tfrac{15}{28}H_{0}(y) - \tfrac{37}{168} + x\Big[- \tfrac{5}{21}H_{0,0,0}(x) + \tfrac{16}{21}H_{0,0}(x)H_{0}(y) - \tfrac{5}{21}H_{0}(x)H_{0,0}(y) \notag\\
  &\quad - \tfrac{85}{63}H_{0,0,0}(y) - \tfrac{5}{9}H_{-1,0,0}(y) - \tfrac{79}{144}\pi^2H_{0}(x) + \tfrac{1025}{432}\pi^2H_{0}(y) - \tfrac{5}{18}\pi^2H_{-1}(y) - \tfrac{25729}{3024}\zeta_3 + \tfrac{5}{18}H_{0,0}(x) \notag\\
  &\quad - \tfrac{37}{18}H_{0}(x)H_{0}(y) + \tfrac{13}{2}H_{0,0}(y) + \tfrac{281}{288}\pi^2 - \tfrac{205}{84}H_{0}(x) + \tfrac{2339}{252}H_{0}(y) + \tfrac{6241}{504}\Big] + y\Big[- \tfrac{85}{63}H_{0,0,0}(x) \notag\\
  &\quad - \tfrac{5}{9}H_{-1,0,0}(x) - \tfrac{5}{21}H_{0,0}(x)H_{0}(y) + \tfrac{16}{21}H_{0}(x)H_{0,0}(y) - \tfrac{5}{21}H_{0,0,0}(y) + \tfrac{1025}{432}\pi^2H_{0}(x) - \tfrac{5}{18}\pi^2H_{-1}(x) \notag\\
  &\quad - \tfrac{79}{144}\pi^2H_{0}(y) - \tfrac{25729}{3024}\zeta_3 + \tfrac{13}{2}H_{0,0}(x) - \tfrac{37}{18}H_{0}(x)H_{0}(y) + \tfrac{5}{18}H_{0,0}(y) + \tfrac{281}{288}\pi^2 + \tfrac{2339}{252}H_{0}(x) \notag\\
  &\quad - \tfrac{205}{84}H_{0}(y) + \tfrac{6241}{504}\Big] + xy\Big[\tfrac{1}{3}H_{0,0}(x) + \tfrac{1}{3}H_{0}(x)H_{0}(y) + \tfrac{1}{3}H_{0,0}(y) + \tfrac{8}{9}\pi^2 - \tfrac{41}{42}H_{0}(x) - \tfrac{41}{42}H_{0}(y) \notag\\
  &\quad + \tfrac{403}{126}\Big]\,, \notag\\
  xy\,d^{(4)}_{-4} &= - \tfrac{152}{63}H_{0,0,0,0}(x) + \tfrac{5}{2}H_{0,-1,0,0}(x) - \tfrac{3}{2}H_{-1,-1,0,0}(x) + \tfrac{16}{63}H_{0,0,0}(x)H_{0}(y) + \tfrac{3}{2}H_{-1,0,0}(x)H_{0}(y) \notag\\
  &\quad + \tfrac{44}{63}H_{0,0}(x)H_{0,0}(y) - \tfrac{17}{126}H_{0}(x)H_{0,0,0}(y) + \tfrac{3}{2}H_{0}(x)H_{-1,0,0}(y) + \tfrac{1}{18}H_{0,-1,0,0}(x/y) \notag\\
  &\quad + \tfrac{4}{9}H_{-1,0,0,0}(x/y) + \tfrac{5}{18}H_{-1,-1,0,0}(x/y) + \tfrac{7}{18}H_{-1,0,0}(x/y)H_{0}(y) - \tfrac{103}{63}H_{0,0,0,0}(y) + \tfrac{5}{2}H_{0,-1,0,0}(y) \notag\\
  &\quad - \tfrac{3}{2}H_{-1,-1,0,0}(y) - \tfrac{103}{54}\pi^2H_{0,0}(x) + \tfrac{5}{4}\pi^2H_{0,-1}(x) + \tfrac{1}{4}\pi^2H_{-1,0}(x) - \tfrac{3}{4}\pi^2H_{-1,-1}(x) \notag\\
  &\quad - \tfrac{43}{27}\pi^2H_{0}(x)H_{0}(y) + \tfrac{3}{4}\pi^2H_{0}(x)H_{-1}(y) + \tfrac{3}{4}\pi^2H_{-1}(x)H_{0}(y) + \tfrac{1}{36}\pi^2H_{0,-1}(x/y) + \tfrac{1}{36}\pi^2H_{-1,0}(x/y) \notag\\
  &\quad + \tfrac{5}{36}\pi^2H_{-1,-1}(x/y) + \tfrac{7}{36}\pi^2H_{-1}(x/y)H_{0}(y) - \tfrac{185}{108}\pi^2H_{0,0}(y) + \tfrac{5}{4}\pi^2H_{0,-1}(y) + \tfrac{1}{4}\pi^2H_{-1,0}(y) \notag\\
  &\quad - \tfrac{3}{4}\pi^2H_{-1,-1}(y) + \tfrac{445}{189}\zeta_3H_{0}(x) - \tfrac{3}{2}\zeta_3H_{-1}(x) + \tfrac{5}{18}\zeta_3H_{-1}(x/y) + \tfrac{995}{378}\zeta_3H_{0}(y) - \tfrac{3}{2}\zeta_3H_{-1}(y) \notag\\
  &\quad - \tfrac{17497}{30240}\pi^4 - \tfrac{74}{21}H_{0,0,0}(x) + 4H_{-1,0,0}(x) - \tfrac{74}{21}H_{0,0}(x)H_{0}(y) - \tfrac{74}{21}H_{0}(x)H_{0,0}(y) - \tfrac{74}{21}H_{0,0,0}(y) \notag\\
  &\quad + 4H_{-1,0,0}(y) - \tfrac{89}{72}\pi^2H_{0}(x) + 2\pi^2H_{-1}(x) - \tfrac{89}{72}\pi^2H_{0}(y) + 2\pi^2H_{-1}(y) - \tfrac{113}{63}\zeta_3 - \tfrac{15}{7}H_{0,0}(x) \notag\\
  &\quad - \tfrac{15}{7}H_{0}(x)H_{0}(y) - \tfrac{15}{7}H_{0,0}(y) - \tfrac{73}{144}\pi^2 + \tfrac{37}{42}H_{0}(x) + \tfrac{37}{42}H_{0}(y) + \tfrac{25}{4} + x\Big[\tfrac{104}{21}H_{0,0,0,0}(x) \notag\\
  &\quad - \tfrac{3}{2}H_{-1,-1,0,0}(x) - \tfrac{64}{21}H_{0,0,0}(x)H_{0}(y) + \tfrac{3}{2}H_{-1,0,0}(x)H_{0}(y) - \tfrac{23}{42}H_{0,0}(x)H_{0,0}(y) \notag\\
  &\quad + \tfrac{103}{42}H_{0}(x)H_{0,0,0}(y) - \tfrac{5}{2}H_{0,-1,0,0}(x/y) - \tfrac{1973}{126}H_{0,0,0,0}(y) + \tfrac{77}{9}H_{0,-1,0,0}(y) + \tfrac{40}{9}H_{-1,0,0,0}(y) \notag\\
  &\quad - \tfrac{5}{9}H_{-1,-1,0,0}(y) + \tfrac{67}{36}\pi^2H_{0,0}(x) + \tfrac{1}{4}\pi^2H_{-1,0}(x) - \tfrac{3}{4}\pi^2H_{-1,-1}(x) + \tfrac{11}{18}\pi^2H_{0}(x)H_{0}(y) \notag\\
  &\quad + \tfrac{3}{4}\pi^2H_{-1}(x)H_{0}(y) - \tfrac{5}{4}\pi^2H_{0,-1}(x/y) - \tfrac{365}{36}\pi^2H_{0,0}(y) + \tfrac{77}{18}\pi^2H_{0,-1}(y) + \tfrac{5}{6}\pi^2H_{-1,0}(y) \notag\\
  &\quad - \tfrac{5}{18}\pi^2H_{-1,-1}(y) - \tfrac{641}{126}\zeta_3H_{0}(x) - \tfrac{3}{2}\zeta_3H_{-1}(x) + \tfrac{4985}{189}\zeta_3H_{0}(y) - \tfrac{5}{9}\zeta_3H_{-1}(y) - \tfrac{63101}{90720}\pi^4 \notag\\
  &\quad - \tfrac{94}{9}H_{0,0,0}(x) + 4H_{-1,0,0}(x) + \tfrac{38}{9}H_{0,0}(x)H_{0}(y) + \tfrac{26}{9}H_{0}(x)H_{0,0}(y) + \tfrac{16}{3}H_{-1,0,0}(x/y) - \tfrac{62}{3}H_{0,0,0}(y) \notag\\
  &\quad + 4H_{-1,0,0}(y) - \tfrac{913}{216}\pi^2H_{0}(x) + 2\pi^2H_{-1}(x) + \tfrac{8}{3}\pi^2H_{-1}(x/y) - \tfrac{89}{72}\pi^2H_{0}(y) + 2\pi^2H_{-1}(y) + \tfrac{359}{36}\zeta_3 \notag\\
  &\quad + \tfrac{3}{7}H_{0,0}(x) + \tfrac{205}{21}H_{0}(x)H_{0}(y) - \tfrac{2339}{63}H_{0,0}(y) - \tfrac{1639}{432}\pi^2 + \tfrac{407}{42}H_{0}(x) - \tfrac{6241}{126}H_{0}(y) - \tfrac{2213}{36}\Big] \notag\\
  &\quad + y\Big[- \tfrac{1144}{63}H_{0,0,0,0}(x) + \tfrac{77}{9}H_{0,-1,0,0}(x) + \tfrac{40}{9}H_{-1,0,0,0}(x) - \tfrac{5}{9}H_{-1,-1,0,0}(x) + \tfrac{104}{21}H_{0,0,0}(x)H_{0}(y) \notag\\
  &\quad - \tfrac{64}{21}H_{0,0}(x)H_{0,0}(y) - \tfrac{23}{42}H_{0}(x)H_{0,0,0}(y) + \tfrac{3}{2}H_{0}(x)H_{-1,0,0}(y) + \tfrac{5}{2}H_{0,-1,0,0}(x/y) + \tfrac{103}{42}H_{0,0,0,0}(y) \notag\\
  &\quad - \tfrac{3}{2}H_{-1,-1,0,0}(y) - \tfrac{205}{18}\pi^2H_{0,0}(x) + \tfrac{77}{18}\pi^2H_{0,-1}(x) + \tfrac{5}{6}\pi^2H_{-1,0}(x) - \tfrac{5}{18}\pi^2H_{-1,-1}(x) \notag\\
  &\quad + \tfrac{67}{36}\pi^2H_{0}(x)H_{0}(y) + \tfrac{3}{4}\pi^2H_{0}(x)H_{-1}(y) + \tfrac{5}{4}\pi^2H_{0,-1}(x/y) + \tfrac{11}{18}\pi^2H_{0,0}(y) + \tfrac{1}{4}\pi^2H_{-1,0}(y) \notag\\
  &\quad - \tfrac{3}{4}\pi^2H_{-1,-1}(y) + \tfrac{4985}{189}\zeta_3H_{0}(x) - \tfrac{5}{9}\zeta_3H_{-1}(x) - \tfrac{641}{126}\zeta_3H_{0}(y) - \tfrac{3}{2}\zeta_3H_{-1}(y) - \tfrac{95231}{90720}\pi^4 - 26H_{0,0,0}(x) \notag\\
  &\quad + 4H_{-1,0,0}(x) + \tfrac{74}{9}H_{0,0}(x)H_{0}(y) - \tfrac{10}{9}H_{0}(x)H_{0,0}(y) + \tfrac{16}{3}H_{-1,0,0}(x/y) - \tfrac{46}{9}H_{0,0,0}(y) + 4H_{-1,0,0}(y) \notag\\
  &\quad - \tfrac{281}{72}\pi^2H_{0}(x) + 2\pi^2H_{-1}(x) + \tfrac{8}{3}\pi^2H_{-1}(x/y) - \tfrac{337}{216}\pi^2H_{0}(y) + 2\pi^2H_{-1}(y) + \tfrac{359}{36}\zeta_3 - \tfrac{2339}{63}H_{0,0}(x) \notag\\
  &\quad + \tfrac{205}{21}H_{0}(x)H_{0}(y) + \tfrac{3}{7}H_{0,0}(y) - \tfrac{1639}{432}\pi^2 - \tfrac{6241}{126}H_{0}(x) + \tfrac{407}{42}H_{0}(y) - \tfrac{2213}{36}\Big] + xy\Big[- \tfrac{16}{3}H_{0,0,0}(x) \notag\\
  &\quad + 6H_{-1,0,0}(x) + \tfrac{2}{3}H_{0,0}(x)H_{0}(y) + \tfrac{2}{3}H_{0}(x)H_{0,0}(y) - \tfrac{16}{3}H_{0,0,0}(y) + 6H_{-1,0,0}(y) - \tfrac{11}{9}\pi^2H_{0}(x) \notag\\
  &\quad + 3\pi^2H_{-1}(x) - \tfrac{11}{9}\pi^2H_{0}(y) + 3\pi^2H_{-1}(y) - \tfrac{35}{18}\zeta_3 - \tfrac{37}{21}H_{0,0}(x) - \tfrac{37}{21}H_{0}(x)H_{0}(y) - \tfrac{37}{21}H_{0,0}(y) \notag\\
  &\quad - \tfrac{239}{54}\pi^2 + \tfrac{121}{21}H_{0}(x) + \tfrac{121}{21}H_{0}(y) - \tfrac{1229}{63}\Big]\,. \label{eq:D4}
\end{align}

\noindent Through $\varepsilon^{4-2\ell}$, as given above, the coefficients contain only harmonic polylogarithms of $x$, $y$ and $x/y$. From $\varepsilon^{5-2\ell}$ on, the alphabet of the $x$-dependence grows to $\{0,-1,-y,-1-y\}$, so the higher orders also contain two-variable multiple polylogarithms.

\end{widetext}

\end{document}